\documentclass[twocolumn,aps,prd,10pt,superscriptaddress,longbibliography]{revtex4-1}
\usepackage{amsmath}
\usepackage{bm}

\usepackage{graphicx}
\usepackage{multirow}
\usepackage{gensymb}
\usepackage{soul}
\usepackage[normalem]{ulem}
\usepackage[usenames,dvipsnames,svgnames]{xcolor}
\usepackage[acronym]{glossaries}
\makeglossaries
\newacronym{nep}{NEP}{neural evolution potential}
\newacronym{dft}{DFT}{density functional theory}
\newacronym{pbe}{PBE}{Perdew-Burke-Ernzerhof}
\newacronym{gap}{GAP}{gaussian approximation potential}
\newacronym{md}{MD}{molecular dynamics}
\newacronym{hnemd}{HNEMD}{homogeneous nonequilibrium molecular dynamics}
\newacronym{nemd}{NEMD}{nonequilibrium molecular dynamics}
\newacronym{snes}{SNES}{separable natural evolution strategy}
\newacronym{sm}{SM}{supplemental material}

\usepackage{hyperref}
\hypersetup{
pdfnewwindow=true,      
colorlinks=true,        
linkcolor=Blue,         
citecolor=Blue,         
filecolor=Blue,         
urlcolor=Blue           
}

\begin{document}

\title{Density dependence of thermal conductivity in nanoporous and amorphous carbon with machine-learned molecular dynamics}

\author{Yanzhou Wang}
\email{yanzhowang@gmail.com}
\affiliation{QTF Center of Excellence, Department of Applied Physics, Aalto University, FIN-00076 Aalto, Espoo, Finland}

\author{Zheyong Fan}
\affiliation{College of Physical Science and Technology, Bohai University, Jinzhou, 121013, China}

\author{Ping Qian}
\email{qianping@ustb.edu.cn}
\affiliation{Beijing Advanced Innovation Center for Materials Genome Engineering, Department of Physics, University of Science and Technology Beijing, Beijing 100083, China}

\author{Miguel A. Caro}
\affiliation{Department of Chemistry and Materials Science, Aalto University, Kemistintie 1, 02150 Espoo, Finland}

\author{Tapio Ala-Nissila}
\email{tapio.ala-nissila@aalto.fi}
\affiliation{QTF Center of Excellence, Department of Applied Physics, Aalto University, FIN-00076 Aalto, Espoo, Finland}
\affiliation{Interdisciplinary Centre for Mathematical Modelling and Department of Mathematical Sciences, Loughborough University, Loughborough, Leicestershire LE11 3TU, United Kingdom}

\date{\today}

\begin{abstract}
Disordered forms of carbon are an important class of materials for applications such as thermal management. However, a comprehensive theoretical understanding of the structural dependence of thermal transport and the underlying microscopic mechanisms is lacking. Here we study the structure-dependent thermal conductivity of disordered carbon by employing molecular dynamics (MD) simulations driven by a machine-learned interatomic potential based on the efficient neuroevolution potential approach. Using large-scale MD simulations, we generate realistic nanoporous carbon (NP-C) samples with density varying from $0.3$ to $1.5$ g cm$^{-3}$ dominated by sp$^2$ motifs, and amorphous carbon (a-C) samples with density varying from $1.5$ to $3.5$ g cm$^{-3}$ exhibiting mixed sp$^2$ and sp$^3$ motifs.  Structural properties including short- and medium-range order are characterized by atomic coordination, pair correlation function, angular distribution function and structure factor. Using the homogeneous nonequilibrium MD method and the associated quantum-statistical correction scheme, we predict a linear and a superlinear density dependence of thermal conductivity for NP-C and a-C, respectively, in good agreement with relevant experiments. The distinct density dependences are attributed to the different impacts of the sp$^2$ and sp$^3$ motifs on the spectral heat capacity, vibrational mean free paths and group velocity. We additionally highlight the significant role of structural order in regulating the thermal conductivity of disordered carbon.
\end{abstract}

\maketitle

\section{Introduction}

Due to its tunability between sp$^2$-bonded graphite (or graphene in 2D) and sp$^3$-bonded diamond, the structurally disordered phases of carbon possess various interesting characteristics~\cite{1999_Lifshitz}. Both flexible sp$^3$/sp$^2$ content ratios and variable mass density enable a wide range of chemical and physical properties in disordered carbon~\cite{2008_balandin_apl,2002_robertson}. It can be approximately divided into two classes. The first displays an abundance of sp$^2$ bonds (typically over $95\%$ content)~\cite{2021_Uskokovic} and has a density typically below that of graphite (about $2$~{g cm$^{-3}$}). Structurally, it comprises curved graphene fragments assembled into $3$D networks~\cite{2019_prl_marks,2022_Wang_cm_NPC}, exemplified by glassy carbon~\cite{2019_shiell_jncs,2021_shiell_jncs,2021_Uskokovic}, nanoporous carbon~\cite{2000_jun_npc,2021_Wang_pccp_DGN, 2022_Wang_cm_NPC} and carbon foams~\cite{2020_chithra_carbonFoam,Chithra2018carbon-foam-newspaper,Raji2023carbon-foam-banana}. The second manifests itself as a mixture of sp$^2$ and sp$^3$ bonds. The sp$^3$ bond fraction varies from $10\%$ to $90\%$ with the density increasing from $2$~{g cm$^{-3}$} (below that of graphite) up to $3.5$~{g cm$^{-3}$} (similar to that of diamond)~\cite{2017_Deringer_prb_GAP17,2021_Heikki_prb_C60,1996_Schwan_a-Csp3,1993_fallon_a-Csp3,2000_ferrari}, as seen in amorphous carbon (a-C) and \textit{tetrahedral} a-C (ta-C), respectively~\cite{1999_Lifshitz,2008_balandin_apl,2002_robertson,2021_scott_irradiatteda-C,2023_Pan_nature_a-C,2021_Shang_nature_a-C}. A more common term, especially in engineering, is ``diamond-like'' carbon (DLC). DLC is a more generic way to refer to amorphous carbon with high sp$^3$ content that may contain chemical impurities, in particular non-negligible amounts of hydrogen~\cite{2002_robertson}.  For convenience, we refer to the first class as nanoporous carbon (NP-C) and the second as a-C from here on. Due to its structural flexibility, disordered carbon finds various applications in energy storage and conversion~\cite{2021_wang_battery,2020_Shao_energyStorage,2021_luo_supercapacitor,Laurila2017_hybrid-carbon}, protective coatings~\cite{2001_robertson_aCCoating,2003_veerasamy_a-CCoating}, wearable electronics~\cite{2019_wang_a-CWeaableElectronics,2023_kim_a-CWeaableElectronics} and wide-bandgap semiconductors~\cite{2006_Tibrewala_a-CBandgap,2023_Abbas_a-Cbandgap,2023_kim_a-CWeaableElectronics}, among others. Notably, many of these applications involve heat management, and thus a comprehensive understanding of heat transport in disordered carbon is of fundamental and practical significance.

To this end, thermal transport properties of disordered carbon have been extensively investigated in recent years. Unlike most other amorphous glass-like materials, the experimentally measured thermal conductivity $\kappa$ varies widely, ranging from $0.1$~\cite{2020_chithra_carbonFoam,Chithra2018carbon-foam-newspaper,Raji2023carbon-foam-banana} to $30$ W m$^{-1}$ K$^{-1}$~\cite{2021_Shang_nature_a-C}. This large span of $\kappa$ values is associated with the following factors. (i) Wide range of mass densities ($\rho$). For instance, on the low-density end, Chithra \textit{et al.}~\cite{2020_chithra_carbonFoam} experimentally prepared ultralight carbon foam samples with densities of $0.15-0.35$ g cm$^{-3}$ by filtering and carbonizing a mixture of sawdust and sucrose, and measured relatively small $\kappa$ values ranging from $0.12$ to $0.2$ W m$^{-1}$ K$^{-1}$. A high-density example, hydrogen-free ta-C films with $2.6-3.3$ g cm$^{-3}$, were experimentally reported to have larger $\kappa$ values in the range of $1.4-3.5$ W m$^{-1}$ K$^{-1}$~\cite{2006_Shamsa_a-CKappa}. Positive correlation between $\kappa$ and $\rho$ is expected to be attributed to the lattice heat capacity, which is proportional to density. (ii) High fraction of sp$^3$ in a-C carbon. Morath \textit{et al.} \cite{1994_morath_a-C-kappa}, Bullen \textit{et al.}~\cite{2000_bullen_a-CKappa} and Shamsa \textit{et al.}~\cite{2006_Shamsa_a-CKappa} all have experimentally concluded that $\kappa$ of a-C scales up with its sp$^3$ content. For example, ta-C films with high sp$^3$ fraction can achieve a $\kappa$ in the range of $5.2-9.7$ W m$^{-1}$ K$^{-1}$~\cite{1994_morath_a-C-kappa}, which is much higher than theoretical predictions and experimental measurements of other amorphous materials. This is theoretically attributed to the sp$^3$ motifs enhancing the relative contribution to $\kappa$ due to propagons (phonon-like vibrational modes) in the low-frequency domain, which significantly improves their diffusivities (the ratio of $\kappa$ to heat capacity), thereby increasing $\kappa$~\cite{2022_npjcm_giri_CN-kappa}. (iii) High medium-range order of the a-C structure. Shang \textit{et al.}~\cite{2021_Shang_nature_a-C} found that more ordered a-C leads to a significant increase in $\kappa$, and samples containing a small amount of nanocrystalline clusters in their matrices show the highest $\kappa=30$ W m$^{-1}$ K$^{-1}$. They argued that the high $\kappa$ is due to the presence of local medium-range order, which promotes the presence of propagons by remarkably elongating their mean free paths (MFPs)~\cite{2021_Shang_nature_a-C}.

Despite considerable efforts to understand thermal conductivity of disordered carbon both experimentally and theoretically, some questions remain open. It has been shown that there is a positive $\kappa \propto \rho$ correlation between $\kappa$ and $\rho$ for disordered carbon as expected from the heat capacity argument mentioned above. An interesting question then pertains to elucidating the precise dependence of $\kappa$ on the density and the microscopic structural characteristics of carbon materials.

There are currently several well-developed theoretical frameworks and computational methods for thermal transport in disordered materials, including the early lattice dynamics-based Allen-Feldman (AF) formulation~\cite{1989_prl_Allen-Feldman} in the harmonic approximation, the more recent quasiharmonic Green-Kubo approximation (QHGK)~\cite{2019_Isaeva_QHGK,kaldo}, and the Wigner transport equation (WTE)~\cite{2019_Michele_nature-physics_WTE,2022_Michele_prx_WTE,Enri2023prr_WTE,simoncelli2023glass-kappa,Harper2024prm_a-Al2O3}. Compared with the earlier AF~\cite{1989_prl_Allen-Feldman}, the more advanced QHGK~\cite{2019_Isaeva_QHGK} and WTE~\cite{2019_Michele_nature-physics_WTE,2022_Michele_prx_WTE} methods allow for partial consideration of anharmonicity and disorder. However, these methods encounter issues such as computational inefficiency for large systems and incomplete consideration of anharmonicity. Molecular dynamics (MD) is a favorable alternative for studying thermal transport in disordered materials due to its affordable linear-scaling computational cost with respect to the number of atoms in the system and its straightforward incorporation of full anharmonicity and realistic disorder. However, a reliable potential energy model for accurately describing interatomic interactions is a prerequisite for predictive MD simulations. It has been demonstrated that, compared with empirical potentials, machine-learned potentials (MLPs) are more capable of capturing the complexity of the local atomic environments in disordered amorphous phases~\cite{2017_Deringer_prb_GAP17,2018_Miguel_prl_aC,2020_Miguel_prb_a-C,2020_jcp_Patrick_carbon,2019_carbon_Carlo_review-carbon-potential,2021_Shaidu_npjcm_PANNA-NNP-carbon,2022_Wang-jinjin_deep_carbon,2023_Minaam-Qamar_ACE-carbon,pegolo2024glass-kappa,Li2024_a-Si-kappa,caro2023_a-Si_a-C}. Among the various MLPs, the neuroevolution potential (NEP)~\cite{2021_fan_nep,2022_fan_nep,2022_fan_gpumd,song2024general} exhibits an excellent computational efficiency and has also been proven to be reliable for heat transport applications in various materials~\cite{dong2024jap,Wu2024jcp_correct-kappa,chen2023_water_ice}, particularly in liquid and amorphous materials \cite{2023_xuke_water,2023_wang_a-Si,2023_liangting_a-SiO2,2023_Honggang-Zhang_prb_HfO2-kappa,pegolo2024glass-kappa}.

In this work, using NEP-driven large-scale MD simulations, we aim to unveil the density and structure dependence of thermal transport in disordered carbon. We generate realistic sp$^2$ bond-dominated NP-C samples with densities varying from $0.3$ to $1.5$ g cm$^{-3}$ and sp$^2$/sp$^3$-mixed a-C samples with densities varying from $1.5$ to $3.5$ g cm$^{-3}$. We systematically characterize their short- and medium-range structural characteristics through atomic coordination, pair correlation function, angular distribution function, and structure factor. Combining the homogeneous nonequilibrium MD method~\cite{2019_fan_prb_hnemd} with the associated quantum-statistical correction scheme, we elucidate a linear and a superlinear density dependence of thermal conductivity for NP-C and a-C by analyzing the different impacts of sp$^2$ and sp$^3$ bonds on the spectral heat capacity, vibrational mean free paths, and group velocity.

\section{Methodology}
\subsection{Datasets}
As verified by our previous work~\cite{2022_Wang_cm_NPC}, the dataset published by Deringer \textit{et al.}~\cite{2017_Deringer_prb_GAP17} for training potential energy ML models of disordered carbon is accurate for both a-C modeling and describing NP-C structures~\cite{2022_Wang_cm_NPC}. Therefore, we use the public dataset (of about $4000$ configurations) as the main part of our training set and additionally incorporate $1890$ bilayer graphene, $720$ low-density NP-C and $40$ glassy carbon configurations. These new structures cover various sp$^2$-bonded atomic environments. Specifically, they vary in interlayer spacing, interlayer slip, distorted lattice and atomic displacement from equilibrium. The NP-C structures range in density from $0.1$ to $1.5$ g cm$^{-3}$. Both NP-C and glassy carbons possess a wide variety of defects compared to pristine graphene such as edges, vacancies, energetically favorable pentagon or heptagon motifs and curvature. These new configurations emphasize sp$^2$ binding in NP-C and are a necessary complement to the original dataset~\cite{2017_Deringer_prb_GAP17}, where low-density turbostratic graphitic carbon structures are lacking. In total, our resultant training set comprises $\approx 6700$ configurations. For the test set, we use the public one~\cite{2017_Deringer_prb_GAP17} and keep it unchanged to allow convenient comparison with available \gls{gap} potentials~\cite{2017_Deringer_prb_GAP17,2022_Wang_cm_NPC}.

We recomputed the two datasets using single-point density functional theory (DFT) calculations with VASP~\cite{1994_Blochl_Projector}. The projector augmented wave (PAW) method~\cite{1994_Blochl_Projector,1999_Kress_paw} with the generalized gradient approximation (GGA) as parameterized by Perdew, Burke and Ernzerhof (PBE)~\cite{1997_Perdew_GGA-PBE} was employed to describe the electronic structure. The energy cutoff for the plane wave expansion was set to 550 eV, the density for $k$-point sampling was $0.2$ \AA$^{-1}$ and the energy convergence threshold was $10^{-6}$ eV.

\subsection{Trained NEP model}

\begin{figure}
    \centering   \includegraphics[width=1\columnwidth]{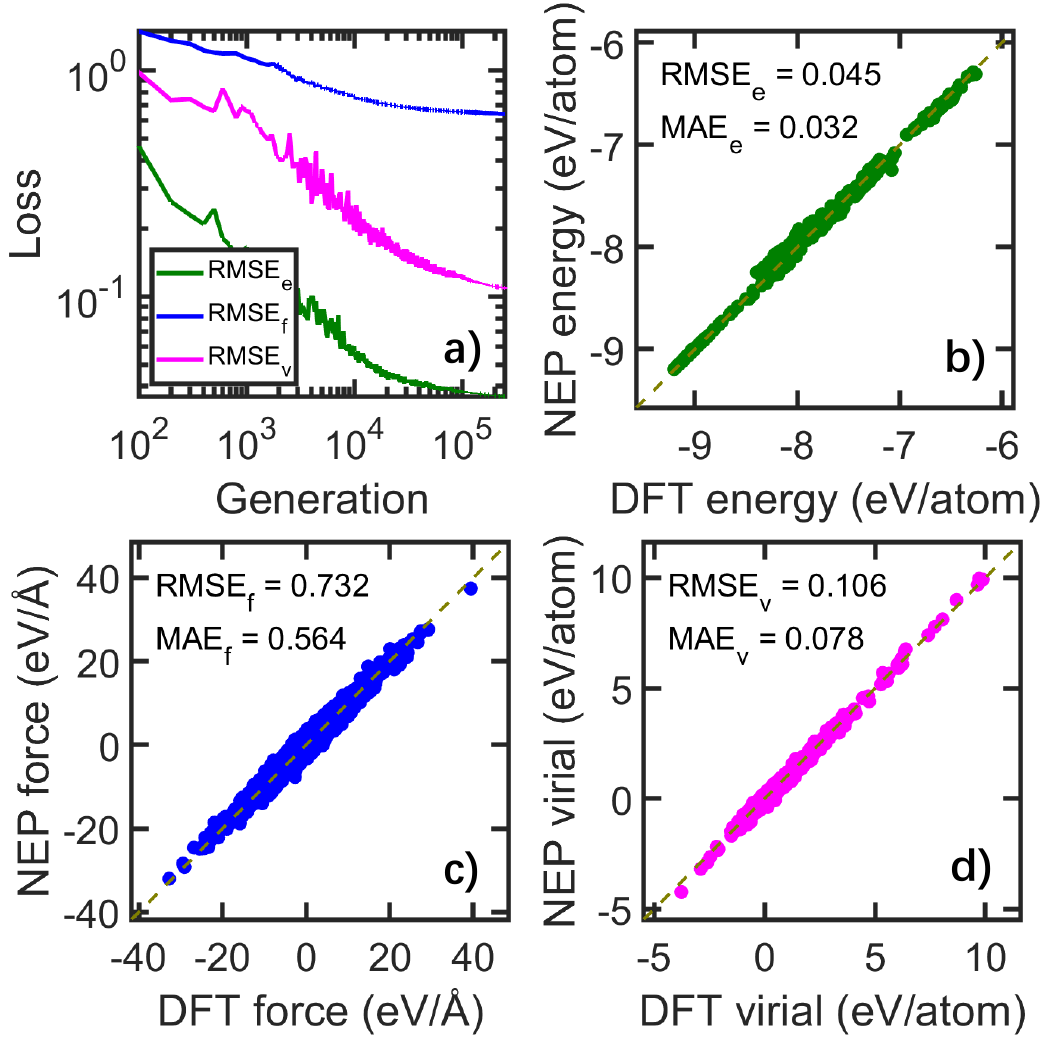}
    \caption{Trained NEP model. (a) Evolution of loss functions of  energy (eV/atom), force (eV/\AA) and virial (eV/atom) RMSEs during training generation for training set. Parities of (b) energy and (c) force and (d) virial  calculated from the NEP model as compared to the DFT-PBE reference data for the test set. Their test RMSEs and MAEs of energy (eV/atom) in (b), force (eV/\AA) in (c) and virial (eV/atom) in (d) are evaluated, respectively.}
    \label{fig:nep-test}
\end{figure}

\begin{table}[]
    \centering
    \setlength{\tabcolsep}{3mm}
    \caption{Test RMSEs of energy (eV/atom)  and force (eV/\AA) by NEP, available GAP17~\cite{2017_Deringer_prb_GAP17} and GAP22~\cite{2022_Wang_cm_NPC}.}
    \begin{tabular}{llll}
    \hline\hline
         & GAP17~\cite{2017_Deringer_prb_GAP17} & GAP22~\cite{2022_Wang_cm_NPC} & NEP \\
         \hline
        Energy RMSE & 0.047 & 0.035 & 0.045 \\
        Force RMSE & 1.14 & 1.04 & 0.73 \\
        \hline\hline      
    \end{tabular}   
    \label{tab:nep}
\end{table}

The \gls{nep} framework~\cite{2021_fan_nep,2022_fan_nep,2022_fan_gpumd} combines a neural network (NN) architecture to represent the potential energy surface and a separable natural evolution strategy (SNES)~\cite{2011_Schaul_snes} for training the model. It uses a feedforward single-hidden-layer NN to map the local atom-environment descriptor of a central atom to its site energy, with the total energy of an extended system given as the sum of the site energies of all of its constituent atoms. The descriptor vector consists of radial and angular components, resembling in spirit Behler-Parrinello symmetry functions~\cite{2011_Behler_jcp} and optimized smooth overlap of atomic positions~\cite{2013_bartok, 2019_Miguel_soap}. The radial descriptor components are a set of radial functions, each of which is formed by a linear combination of Chebyshev polynomials. The angular descriptor components include 3- and 4-body correlations constructed in terms of spherical harmonics. \gls{nep} uses the SNES technique to optimize the free parameters by globally minimizing a loss function that is a weighted sum of the root-mean-square errors (RMSEs) of energy, force and virial stress.

Regarding the hyperparameter settings required for training  the \gls{nep} model, we use a cutoff radius of $4.2$ \AA~ for the radial descriptors and $3.7$ \AA~ for the angular descriptors. These and other hyperparameters for the NEP model training are listed in Note S1 of the Supplementary Material (SM). Figure \ref{fig:nep-test}(a) presents the evolution of RMSEs of energy, force, and virial during the training iteration. Upon convergence, these quantities predicted by the \gls{nep} model on the training set are plotted in Fig. S1 against the DFT-PBE reference data. As shown in Fig.~\ref{fig:nep-test}(b-d), the trained \gls{nep} overall performs well in predictions of energy, force, and virial, with the test errors of $\rm{RMSE_e}=0.045$ (mean absolute error, $\rm{MAE_e}=0.032$) eV/atom, $\rm{RMSE_f}=0.732$ ($\rm{MAE_f}=0.564$) eV/\AA, and $\rm{RMSE_v}=0.106$ ($\rm{MAE_v}=0.078$) eV/atom, respectively. For comparison, Table \ref{tab:nep} lists the energy and force RMSEs of the trained \gls{nep} and the available \gls{gap} models for the same test set. Overall, our NEP model demonstrates intermediate energy accuracy and improved force accuracy as compared to the previous \gls{gap}22~\cite{2022_Wang_cm_NPC}  and \gls{gap}17~\cite{2017_Deringer_prb_GAP17} models. It is worth noting that our NEP model demonstrates higher accuracy for disordered carbon, with force RMSEs of $0.3$~eV/\AA~for NP-C and $0.5$~eV/\AA~for a-C (Fig.~S16), compared to our reported RMSE of $0.73$~eV/\AA~on the test set that includes liquid structures with large forces. Force-fitting errors in MLPs generally act as external perturbations to interatomic forces, leading to an underestimation of thermal conductivity~\cite{Wu2024jcp_correct-kappa}. This effect is particularly pronounced in highly harmonic crystals with high thermal conductivity, such as diamond (see thermal conductivity results in Fig.~S17), but has a negligible impact on disordered materials~\cite{Wu2024jcp_correct-kappa}.

\subsection{Thermal transport}

In the \gls{hnemd} method~\cite{2019_fan_prb_hnemd}, one can induce a heat current by applying an external driving force
\begin{equation}
\label{equation:Fe}
\bm{F}_{i}^{\rm ext} = \bm{F}_{\rm e}\cdot\mathbf{W}_{i}.
\end{equation}
Here, $\bm{F}_{\rm e}$ denotes the driving force parameter (dimension of inverse length) and $\mathbf{W}_{i}$ is the $3\times 3$ virial tensor of atom $i$ defined as~\cite{2021_fan_nep,2015_fan_force-gpumd,2021_prb_Gabourie_Kt}
\begin{equation} 
\label{equation:virial}
\mathbf{W}_{i} = \sum_{j\neq i} \bm{r}_{ij} \otimes \frac{\partial U_j}{\partial \bm{r}_{ji}},
\end{equation}
where the displacement vector is written as $\bm{r}_{ij} \equiv \bm{r}_j - \bm{r}_i$, $\otimes$ denotes a tensor product between two vectors, and $U_j$ is the site energy of atom $j$.

In the linear-response theory, the nonequilibrium ensemble average of the heat current $\bm{J}$ is proportional to $\bm{F}_{\rm e}$:
\begin{equation}
\label{equation:J1}
\langle J^{\alpha} \rangle =  T V \sum_{\beta} \kappa^{\alpha\beta} F_{\rm e}^{\beta},
\end{equation}
where $T$ and $V$ denote system temperature and volume, respectively. We can compute the instant heat current using the definition~\cite{2021_prb_Gabourie_Kt,2015_fan_force-gpumd,2021_fan_nep},
\begin{equation} 
\label{equation:J2}
\bm{J} = \sum_{i} \mathbf{W}_{i} \cdot \bm{v}_{i},
\end{equation}
where $\bm{v}_i$ denotes the velocity of atom $i$. Therefore, combining with the Eq.~\eqref{equation:J1} and Eq.~\eqref{equation:J2}, one can obtain the thermal conductivity tensor $\kappa^{\alpha\beta}$ by sampling the time (MD) average of $\bm{J}$ within steady state generated by the external driving force with a given $\bm{F}_{\rm e}$.

In the HNEMD formalism, one can obtain the spectral thermal conductivity $\kappa(\omega)$  as a function of the vibrational frequency $\omega$, which is derived from the virial-velocity correlation function~\cite{2019_fan_prb_hnemd,2021_prb_Gabourie_Kt}
\begin{equation}
    \bm{K}(t) = \sum_i \langle \mathbf{W}_i(0) \cdot \bm{v}_i(t) \rangle.
\end{equation}
The summation is over the atoms in a system of interest. For simplicity, we will assume to work with a diagonal component of the thermal conductivity tensor and drop the tensorial notations. The spectral thermal conductivity is then obtained via Fourier transform of the virial-velocity correlation function:
\begin{equation}
    \kappa(\omega, T) = \frac{2}{VTF_{\rm e}}\int_{-\infty}^{\infty} \text{d}t \, e^{\text{i}\omega t} K(t),
    \label{eq:kw}
\end{equation}
where we have emphasized the temperature dependence of the thermal conductivity. 
The ratio between the diffusive spectral thermal conductivity and the ballistic spectral thermal conductance $G(\omega)$ defines the spectral mean free path: \cite{2019_fan_prb_hnemd}
\begin{equation}          
      \lambda(\omega,T)=\kappa(\omega,T)/G(\omega).
      \label{eq:mfp}
\end{equation}
The spectral thermal conductance $G(\omega)$ can be obtained by a \gls{nemd} simulation in the ballistic regime and similar Fourier transform of the virial-velocity correlation function:
\begin{equation}
    G(\omega) = \frac{2}{V\Delta T}\int_{-\infty}^{\infty} \text{d}t \, e^{\text{i}\omega t} K(t),
\end{equation}
where $\Delta T$ is the temperature difference between the 
heat source and heat sink in the \gls{nemd} setup.

\subsection{MD computational details}

\begin{figure}
    \centering    
    \includegraphics[width=1\columnwidth]{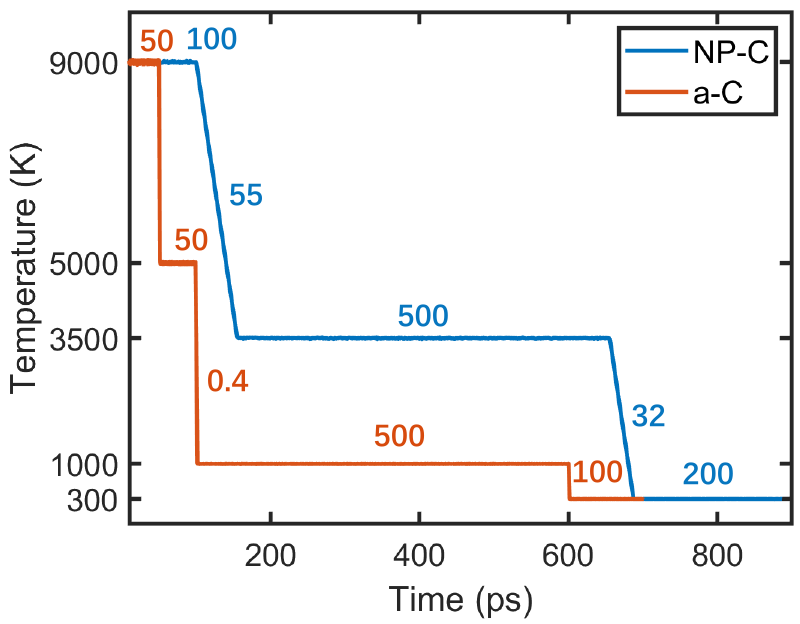}
    \caption{Examples of temperature protocols for $125000$-atom sample generations of $0.5$ g cm$^{-3}$ NP-C and $3.5$ g cm$^{-3}$ a-C in \gls{md} simulation.}
    \label{fig:temperature}
\end{figure}

We generated NP-C and a-C samples using different melt-quench temperature profiles in \gls{md} simulations. During the melting and quenching processes, the constant or varying target temperatures are realized by using the Berendsen thermostat~\cite{1984_berendsen_bdpthermostat} with a time constant of $100$ fs.  A model based on $25\times25\times25$ diamond supercell (containing $125,000$ atoms) was used as the initial structure. For the NP-C preparation, we followed the previous melt-graphitization-quench protocol~\cite{2022_Wang_cm_NPC} but kept the samples at the graphitization stage for a longer time. Specifically, we first heated the samples to a temperature of $T_0=9000$~K, which enabled the atoms to become highly disordered in a short time. After $100$ ps, we rapidly quenched the samples from $T_0$ to the graphitization stage at $T_1^{\text{NP-C}}=3500$ K (slightly below the melting point $T_{\rm m}\approx 4000$~K) over $55$ ps. Compared to our previous work~\cite{2022_Wang_cm_NPC}, where the graphitization duration was $200$ ps, we extended this process to $500$ ps to ensure thorough graphitization. The graphitized NP-C samples were then cooled down to room temperature over $32$ ps and finally annealed for $200$ ps.

For the a-C preparation, liquid carbon samples at $9000$ K were cooled down to a lower temperature (but still above $T_{\rm m}$) at $T_1^{\text{a-C}}=5000$ K and reequilibrated for another $50$ ps. We then performed a rapid melt-quench process from $5000$ to $1000$ K. This is a key stage for disordered a-C preparation, as the quenching rate has a remarkable effect on the ordering of a-C structures. Our tests show that rapid quenching with a rate of $\alpha=10^{15}-10^{16}$ K/s tends to form more realistic a-C, whereas an overly low quenching rate, such as $\alpha=10^{11}-10^{12}$ K/s, leads to crystallization to a certain extent, which will be discussed in the following. We chose $\alpha=10^{16}$ K/s for a-C generation, except for the discussion about the ordering effect on thermal conductivity in Sec.~\ref{sec:order}, where $\alpha$ is varied to monitor its impact on $\kappa$. It took $0.4$ ps to quench from the low-temperature ($T_1^{\text{a-C}}$) melting state to the solid a-C at $T_2^{\text{a-C}}=1000$ K, followed by an annealing procedure for $500$ ps. This long-time relaxation ensures the removal of metastable motifs and residual internal stresses resulting from the rapid quench. Finally, the system was equilibrated at $300$ K for $100$ ps.

Using the aforementioned melt-graphitization-quench and rapid melt-quench protocols, we prepared NP-C and a-C samples, respectively, with a wide range of densities. Specifically, the NP-C samples vary in density as $0.3$, $0.5$, $0.75$, $1$, $1.25$, and $1.5$ g cm$^{-3}$, corresponding to model sizes of $20.3$, $17.1$, $14.9$, $13.6$, $12.6$, and $11.8$ nm, respectively. The a-C samples cover densities of $1.5$, $2$, $2.5$, $3$, and $3.5$ g cm$^{-3}$, with sizes of $11.8$, $10.8$, $10$, $9.4$, and $8.9$ nm, respectively. Figure~\ref{fig:temperature} presents examples of temperature-time profiles for $0.5$ g cm$^{-3}$ NP-C (in blue) and $3.5$ g cm$^{-3}$ a-C (in red) in practical \gls{md} simulations. For the sake of statistical accuracy in the structural and thermal properties of disordered carbon, we performed three independent runs for the preparation of samples at each density.

After generating NP-C and a-C samples, we used the Open Visualization Tool (OVITO)~\cite{2010_stukowski_ovito} to render the structural morphology and compute the atomic coordination of atoms with a cutoff distance of $1.9$ \AA. Structural properties such as the pair correlation function (PCF), angular distribution function (ADF), and static structure factor were calculated using the Interactive Structure Analysis of Amorphous and Crystalline Systems (ISAACS) package~\cite{2010_roux_isaacs}.

For the calculations of thermal transport, we used the NVT ensemble with the Nose-Hoover chain thermostat~\cite{2023_tuckerman_Nose-Hoover-chain} to perform \gls{hnemd} calculations for $6$ ns after $0.2$ ns of equilibration at $300$ K. A system size of 125,000 atoms (corresponding to $8.9$~nm for $3.5$~g~cm$^{-3}$ a-C and $11.8$~nm for $1.5$~g~cm$^{-3}$ NP-C) was sufficient to eliminate finite-size effects on the thermal conductivity of disordered carbon (see Fig.~S2).
 For the \gls{nemd} simulations, we froze some atomic layers with a certain thickness at the two ends of the sample to create fixed boundaries. The two blocks next to the fixed domains were chosen as hot (heat source) and cold (heat sink) baths. The zone of heat flux between two thermal baths were uniformly divided into three groups. The Langevin thermostat~\cite{2007_bussi_langevin} for heat source and heat sink were applied to generate a nonequilibrium heat current. In our \gls{nemd} calculations, the thicknesses of the two regions coupled to the thermal baths were both set to $4.5$ nm for NP-C and $3$ nm for a-C, and the thickness of each group in the heat flux zone was set to $0.5$ nm. To obtain the ballistic thermal conductance (no phonon scattering) as realistically as possible, we set the target temperatures to $T_{\rm h}=25$ K for the heat source and $T_{\rm c}=15$ K for the heat sink, and only considered the contribution of heat flux from the middle group with the thickness of $0.5$ nm. All \gls{md} calculations were run with the \textsc{gpumd} package~\cite{2017_fan_gpumd}. The time step was set to $0.5$ fs (unless stated otherwise) for \gls{md} calculations. 

\section{Results and discussion}
\subsection{Structural characteristics of NP-C and a-C}\label{sec:structure}

\begin{figure*}
    \centering    
    \includegraphics[width=1.8\columnwidth]{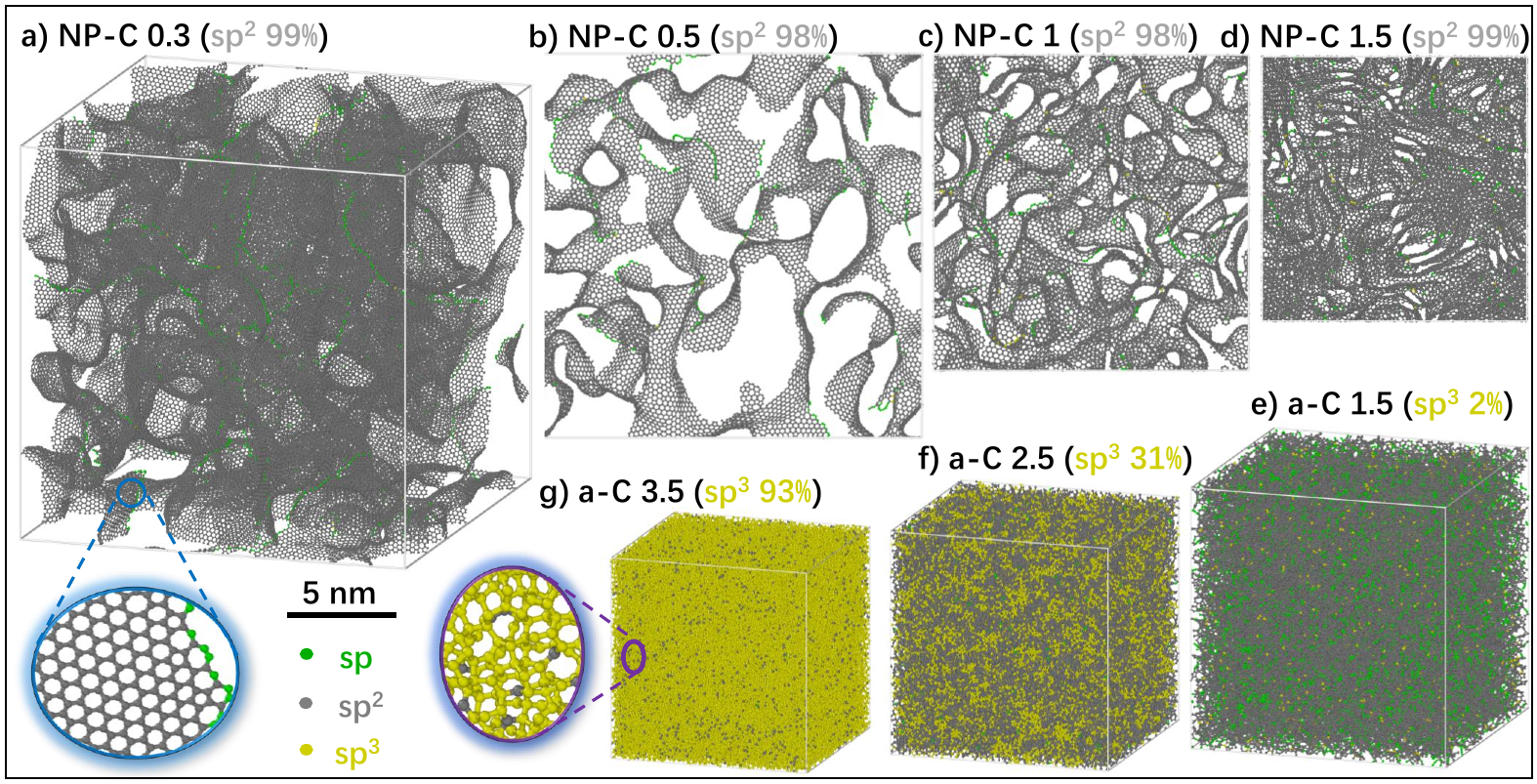}
    \caption{Visualizations of 125000-atom (a-d) NP-C and (e-g) a-C MD snapshots. (a) 3D model with density of  $0.3$ g cm$^{-3}$, and 2-nm-thick slices with densities of (b) $0.5$, (c) $1$ and (d) $1.5$ g cm$^{-3}$ for NP-C samples, as compared with 3D a-C with densities of (e) $1.5$, (f) $2.5$ and (g) $3.5$ g cm$^{-3}$. sp, sp$^2$ and sp$^3$ atoms are rendered in green, gray and blue, respectively. The fractions of sp$^2$ for NP-C and sp$^3$ for a-C are also presented in each panel.}
    \label{fig:morphology}
\end{figure*}

\begin{figure}
    \centering
    \includegraphics[width=1\columnwidth]{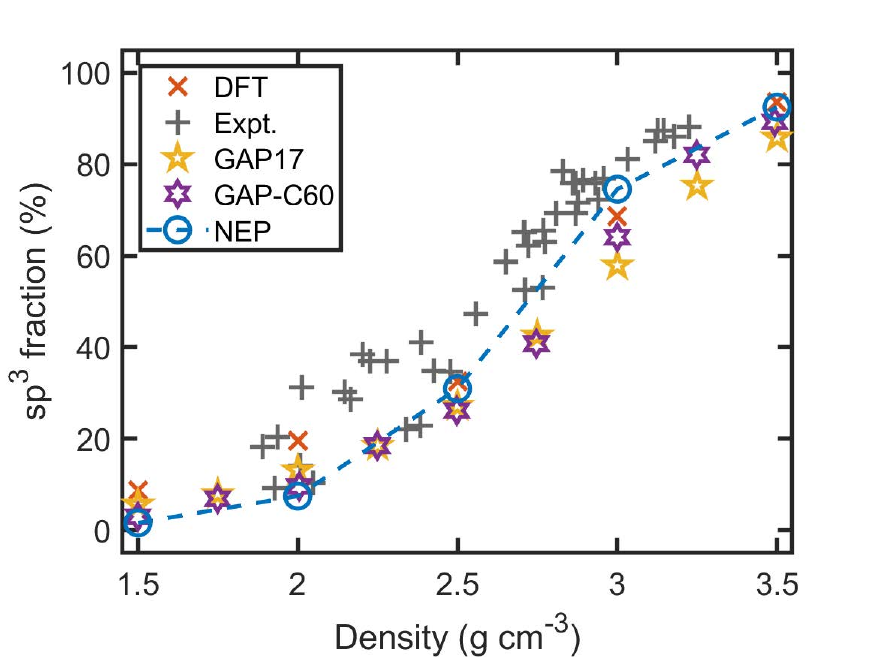}
    \caption{sp$^3$ fraction as a function of mass density for our a-C samples (NEP), as compared to those of DFT calculations~\cite{2017_Deringer_prb_GAP17}, experimental measurements~\cite{1993_fallon_a-Csp3,1996_Schwan_a-Csp3,2000_ferrari}, as well as GAP17~\cite{2017_Deringer_prb_GAP17} and GAP-C60~\cite{2021_Heikki_prb_C60} MD predictions. }
    \label{fig:CN}
\end{figure}

\begin{figure*}
    \centering
    \includegraphics[width=1.8\columnwidth]{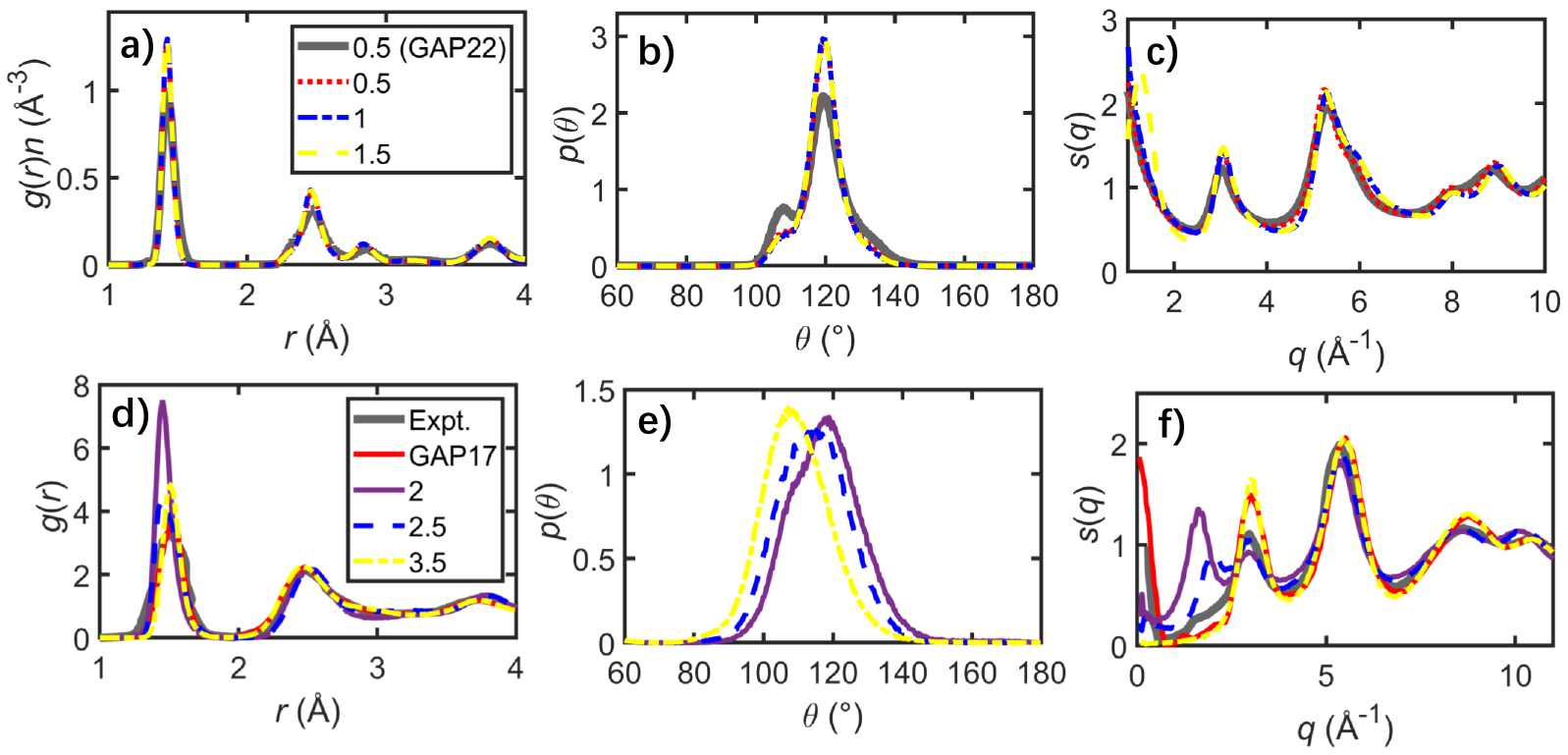}
    \caption{(a, d) Pair correlation function $g(r)n$ ($n$ is the atomic density with units of atoms/\AA$^3$) or $g(r)$, (b, e) angular distribution function $p(\theta)$ and (c, f) structure factor $s(q)$ for the density-varying (a-c) NP-C ($\rho=0.5$, $1$ and $1.5$ g cm$^{-3}$) and (d-f) a-C ($\rho=2$, $2.5$ and $3.5$ g cm$^{-3}$) structures. GAP22 calculation of $g(r)n$ for $0.5$ g cm$^{-3}$ NP-C~\cite{2022_Wang_cm_NPC}, as well as GAP17 deposition simulations~\cite{2018_Miguel_prl_aC} and experimental data~\cite{1995_Gilkes_prb_aC} of both $g(r)$ and $s(q)$ for $\sim 3.5$ g cm$^{-3}$ a-C are presented for comparison.}
    \label{fig:pcf_adf_sq}
\end{figure*}

We present the morphologies of sp$^2$-bonded NP-C  and sp$^2$/sp$^3$-hybridized a-C samples with different densities. As shown in Fig.~\ref{fig:morphology} (a-d), NP-C samples comprise curved graphene fragments assembled into 3D networks~\cite{2019_prl_marks,2022_Wang_cm_NPC}, and display a monotonic decrease in pore size with increasing density. Additionally, the calculation of atomic coordination in Table S1 shows that the sp$^2$-bonded motifs of NP-C dominate with a proportion of at least $98\%$, and this high fraction is essentially independent of its density. These results agree well with our previous \gls{gap}22-driven \gls{md} work~\cite{2022_Wang_cm_NPC}. However, high-density a-C possesses different features in this regard. As seen in Fig.~\ref{fig:morphology} (e-g), the sp$^2$- and sp$^3$-bonded atoms are uniformly distributed in the a-C structures, and the sp$^3$ fraction dramatically increases from $2\%$ to $93\%$ as the density ranges from $1.5$ to $3.5$ g cm$^{-3}$. To correlate the variation in atomic coordination with other structural parameters, we provide the sp$^3$ fractions of a-C as a function of density in Fig.~\ref{fig:CN}, along with available data from DFT~\cite{2017_Deringer_prb_GAP17} and \gls{gap}-\gls{md}~\cite{2017_Deringer_prb_GAP17,2021_Heikki_prb_C60} simulations and experimental measurements~\cite{1993_fallon_a-Csp3,1996_Schwan_a-Csp3,2000_ferrari}. The good agreement indicates the robustness of our \gls{nep} in describing the atomic bonding of a-C structures. Note that MD-based melt-quench simulations of a-C structures are not expected to fully match experimental observables, for which a more sophisticated simulated-deposition protocol (closer to the experimental growth conditions) is needed~\cite{2018_Miguel_prl_aC,2020_Miguel_prb_a-C}. Nevertheless, for the purpose of understanding the impact of a-C structure on thermal conductivity, the melt-quench protocol provides a computationally affordable alternative with structures close enough to the experimental ones.

We further characterized the short- and medium-range structural properties of NP-C and a-C by analyzing the PCF, ADF, and static structure factor. As in our previous work~\cite{2022_Wang_cm_NPC}, we continue to use the product of the PCF $g(r)$ and particle density $n$ to highlight comparisons between NP-C systems of different densities. As seen in Fig.~\ref{fig:pcf_adf_sq}(a), the $g(r)n$ curves for densities $\rho = 0.5$, $1$, and $1.5$ g cm$^{-3}$ almost completely overlap. Additionally, the ADFs in Fig.~\ref{fig:pcf_adf_sq}(b), which complement the analysis of bond angles of atoms in the first neighbor shell, show a similar overlap among systems of different densities. Compared with \gls{gap}22~\cite{2022_Wang_cm_NPC}, our \gls{nep} model predicts slightly sharper distributions for both the PCF and ADF of $0.5$ g cm$^{-3}$ NP-C, in describing the first two peaks of $g(r)n$ at $r_1=1.42$ and $r_2=2.46$ \AA~in Fig.~\ref{fig:pcf_adf_sq}(a), as well as the main $p(\theta)$ peak at $120\degree$ (corresponding to 6-membered rings) in Fig.~\ref{fig:pcf_adf_sq}(b). This suggests a slightly more ordered structural feature in the short range. The small differences in ordering are likely due to the duration of graphitization, with the \gls{nep} run achieving more thorough graphitization for NP-C than \gls{gap}22 ($500$ vs. $200$ ps) during the sample generation at $3500$~K.

Additionally, the more thorough graphitization is also evidenced by the shoulder peaks of the ADF in Fig.~\ref{fig:pcf_adf_sq}(b). Besides the 6-membered rings predominantly making up the graphene fragments in the NP-C structure, 5- and 7-membered rings are energetically favorable defect motifs that usually exist in 3D disordered networks~\cite{2022_Wang_cm_NPC,2019_prl_marks}. These defected rings correspond to the positions of the $p(\theta)$ shoulder peaks at $\theta_1 \approx 108\degree$ (internal angle of pentagon) and $\theta_2 \approx 129\degree$ (heptagon angle), respectively. We see that the two shoulder peaks diminish with more thorough graphitization.

We further examine the medium-range characteristics of NP-C structures by analyzing their structure factors, which is computationally the Fourier transform of $g(r)$, i.e., $s(q)=1+4\pi n\int_0^{\infty}r^2({\sin{(qr)}}/{qr})[g(r)-1]\text{d}r$. As shown by the analyses of the PCF and ADF in Fig.~\ref{fig:pcf_adf_sq}(a-b), the $s(q)$ results for NP-C in Fig.~\ref{fig:pcf_adf_sq}(c) also demonstrate a similar insensitivity to density, with the exception of an additional peak at $q=1.3$ \AA$^{-1}$ for the density of $1.5$ g cm$^{-3}$. This peak is related to local graphite-like pockets (layer-stacking motifs) in the NP-C structure. We observe that the sample with a density of $1.5$ g cm$^{-3}$, which is $\approx 66$\% that of graphite, locally forms interlayer stacking of graphene sheets in the NP-C morphology shown in Fig.~\ref{fig:morphology}(d). Likewise, in our prior analysis of the diffraction pattern, this prepeak in the $1.5$ g cm$^{-3}$ sample has been identified as the $\{002\}$ reflection~\cite{2022_Wang_cm_NPC}. Finally, our calculated $s(q)$ for the $0.5$ g cm$^{-3}$ NP-C agrees well with the results from \gls{gap}22~\cite{2022_Wang_cm_NPC}.

Based on the above analyses of PCFs, ADFs, and structure factors for NP-C samples with varying densities, we clearly conclude that the structural characteristics of the in-plane graphene sheets in NP-C samples are essentially independent of density. This point is significant as it provides the necessary context to develop our understanding of the correlation between thermal conductivity and density in the following sections.

We turn to the structural properties of a-C. The increase of sp$^3$ content with density implies the increasing diamond-like character in sp$^2$/sp$^3$-hybridized a-C structures. In Fig.~\ref{fig:pcf_adf_sq}(d-e), we see that the position of the first peak of $g(r)$ shifts from $r_1=1.45$~\AA~(close to the graphene bond) at $2$ g cm$^{-3}$ to $r_1^{\prime}=1.52$~\AA~(close to the diamond bond) at $3.5$ g cm$^{-3}$, and the position of the $p(\theta)$ peak shifts from the lower-density sp$^2$-bonded hexagon $\theta_1=120\degree$ to the higher-density sp$^3$-bonded tetrahedron $\theta_1^{\prime}=109.5\degree$. The PCF data from experiments~\cite{1995_Gilkes_prb_aC} and deposition simulations~\cite{2018_Miguel_prl_aC} for $\rho\approx3.5$ g cm$^{-3}$ a-C films are provided for comparison. We see that our calculated PCF for $3.5$ g cm$^{-3}$ a-C agrees with these references, but shows slightly more short-range ordering characteristics.

Figure~\ref{fig:pcf_adf_sq}(f) presents the structure factors of a-C structures. For the density of $3.5$ g cm$^{-3}$, our \gls{nep}-derived result is generally consistent with the \gls{gap}17 simulation~\cite{2018_Miguel_prl_aC}, yet it indicates a first peak at $q\approx3$ \AA$^{-1}$ higher than that in the experimental measurements~\cite{1995_Gilkes_prb_aC}. Medium-range ordering is typically associated with the first peak of $s(q)$, and we argue that our a-C sample structurally possesses slightly enhanced ordering in the medium range. Moreover, from Fig.~\ref{fig:pcf_adf_sq}(f) we observe that a-C samples with lower densities develop new prepeaks, and tend to grow earlier and sharper prepeaks. Due to the dominance of sp$^2$-bonded motifs in a-C samples with densities $\rho<2.7$ g cm$^{-3}$, we reason that denser and smaller stacking pockets in a-C as compared to those in $1.5$ g cm$^{-3}$ NP-C are responsible for these prepeaks. Based on the discussion of PCF, ADF and structure factor for a-C, we conclude that our generated a-C samples show the required structural similarity to experimental ones~\cite{1995_Gilkes_prb_aC}, but demonstrate slightly more enhanced ordering characteristics in both short and medium ranges. Structural ordering has a positive effect on thermal transport~\cite{2011_Suarez-Martinez_struc-kappa,2021_Shang_nature_a-C}, which will be systematically discussed in the following Sec.~\ref{sec:order}.

\subsection{Thermal transport of NP-C and a-C}

\subsubsection{Density-dependent thermal conductivity}

\begin{figure*}
    \centering  
    \includegraphics[width=1.8\columnwidth]{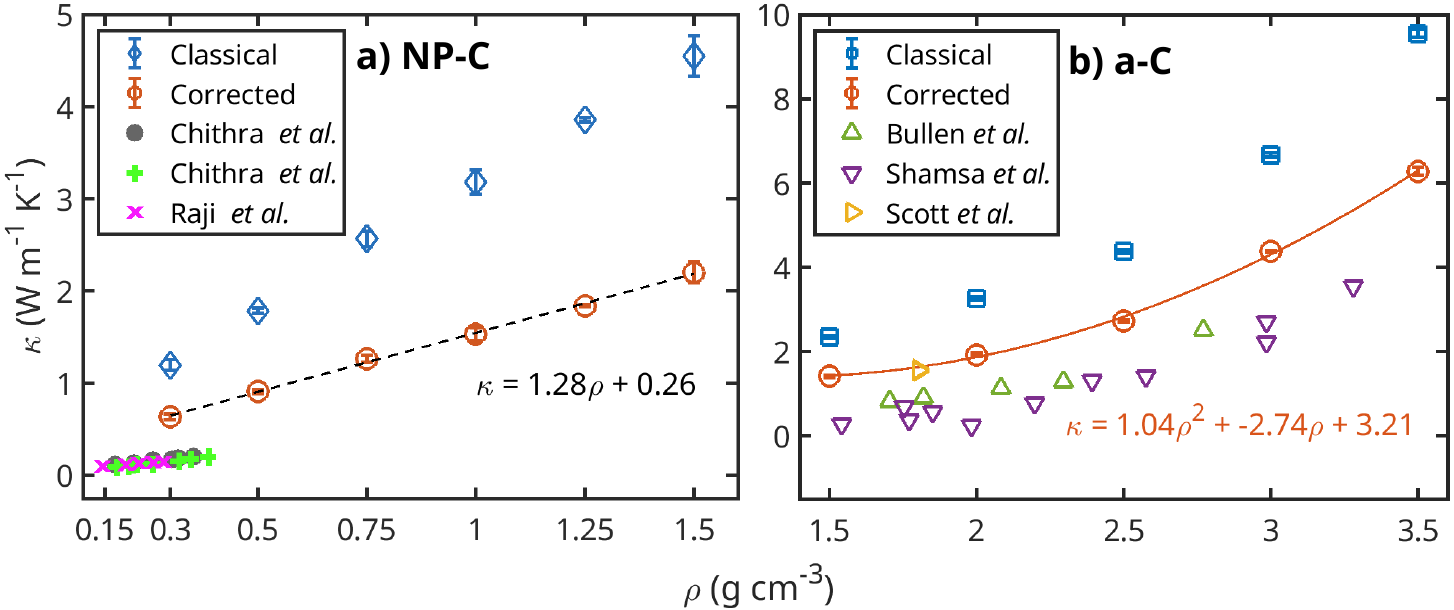}
    \caption{Classical and quantum-corrected thermal conductivity $\kappa$ as a function of density $\rho$ for the (a) NP-C ($\rho=0.3 - 1.5$ g cm$^{-3}$)  and (b) a-C ($\rho=1.5 - 3.5$ g cm$^{-3}$), as compared to experimental values from carbon foam by Chithra \textit{et al.} \cite{2020_chithra_carbonFoam,Chithra2018carbon-foam-newspaper} and Raji \textit{et al.} \cite{Raji2023carbon-foam-banana}, as well as a-C by Bullen \textit{et al.} \cite{2000_bullen_a-CKappa}, Shamsa \textit{et al.} \cite{2006_Shamsa_a-CKappa} and Scott \textit{et al.} \cite{2021_scott_irradiatteda-C}. The quantum-corrected $\kappa(\rho)$ of NP-C and a-C are correlated by linear and high-order polynomial fittings, respectively.}
    \label{fig:k-rho}
\end{figure*}

In this section, we study the thermal transport properties of disordered carbons with varying densities using \gls{hnemd} calculations. To assess the degree of ergodicity and quantify error statistics in the calculation of thermal properties, we performed three independent runs at each density and set the threshold of running time to $6$ ns for each density at $300$ K. The plots of the cumulative average of thermal conductivity $\kappa$ versus production time for all densities of NP-C ($\rho = 0.3-1.5$ g cm$^{-3}$) and a-C ($\rho = 1.5-3.5$ g cm$^{-3}$) are provided in Figs. S3 and S4, respectively. The smooth convergence of averaged $\kappa$ with time for all \gls{md} runs validates the choice of input parameter of driving force $F_{\rm{e}}$ in \gls{hnemd} calculations. We used these time-converged $\kappa$ values to plot the classical $\kappa (\rho)$ correlation in blue squares in Fig.~\ref{fig:k-rho}(a) for NP-C and Fig.~\ref{fig:k-rho}(b) for a-C, along with representative experimental data from Chithra \textit{et al.}~\cite{2020_chithra_carbonFoam,Chithra2018carbon-foam-newspaper} and Raji \textit{et al.}~\cite{Raji2023carbon-foam-banana} for NP-C, as well as Bullen \textit{et al.}~\cite{2000_bullen_a-CKappa}, Shamsa \textit{et al.}\cite{2006_Shamsa_a-CKappa}, and Scott \textit{et al.}~\cite{2021_scott_irradiatteda-C} for a-C. Considering the assumption of classical statistics that all vibrational modes are fully excited and the high Debye temperature $\Theta_{\rm D} \approx 2300$ K for disordered carbon, we see substantial overestimates of $\kappa$ in both NP-C and a-C (see blue squares in Fig.~\ref{fig:k-rho}) compared to experiments. We attribute the overshooting mostly to the missing quantum-statistical effects in the \gls{md} simulations, which we address below.

There is a feasible technique for applying quantum statistical corrections to spectral thermal conductivity $\kappa (\omega)$ in Eq.~\eqref{eq:kw} in \gls{hnemd} calculations. This is achieved by multiplying it by the ratio between quantum and classical modal heat capacities~\cite{2016_lv_qc,2016_saaskilahti_qc22,2017_fan_qc3} to obtain quantum-corrected spectral thermal conductivity:
\begin{equation}
\kappa^{\rm q}(\omega) = \kappa(\omega)\frac{x^2e^x}{(e^x-1)^2},
\label{equation:qc_k_omega}
\end{equation}
where $x=\hbar\omega/k_{\rm B}T$, $\hbar$ denotes the reduced Planck constant, and $k_{\rm B}$ is the Boltzmann constant. We emphasize that the validity of the quantum-correction approximation is confined to disordered systems where the population of vibrations has negligible effects on elastic scattering processes due to their very short lifetimes or mean free paths. Recently, this correction method has been further validated in $\kappa$ calculations for disordered silicon~\cite{2023_wang_a-Si}, silica~\cite{2023_liangting_a-SiO2}, and even liquid water~\cite{2023_xuke_water}.

Upon integrating $\kappa^q(\omega)$ for all densities of disordered samples at $T=300$ K in Fig.~\ref{fig:kw}, we present the quantum-corrected $\kappa$ in red circles in Fig.~\ref{fig:k-rho}. Compared with the classical $\kappa$, the quantum-corrected $\kappa^q$ are reduced by about $50\%$ for NP-C and $30\%-40\%$ for a-C, thus bringing them much closer to the experimental values. Furthermore, to gain deeper insights into the $\kappa(\rho)$ dependence, we performed polynomial fittings on $\kappa^q$. This clearly shows a linear $\kappa^{\rm{NP-C}}(\rho)\propto \rho$ correlation for NP-C in Fig.~\ref{fig:k-rho}(a), but a \textit{superlinear} $\kappa^{\rm{a-C}}(\rho)$ dependence on $\rho$ for a-C in Fig.~\ref{fig:k-rho}(b). 

Based on the knowledge of structure-property relationships, we reason that the difference in $\kappa (\rho)$ dependencies between NP-C and a-C primarily originates from their distinct structures\textemdash{}specifically, the density-independent nature of the sp$^2$-dominated graphene-like fragments in NP-C against the density-dependent sp$^3/(\rm{sp}^2+\rm{sp}^3)$ ratio in a-C, as discussed in Sec.~\ref{sec:structure}. For comparison, we consider experimental measurements of disordered samples with similar structural characteristics. Chithra \textit{et al.}~\cite{Chithra2018carbon-foam-newspaper,2020_chithra_carbonFoam} and Raji \textit{et al.}~\cite{Raji2023carbon-foam-banana} experimentally prepared micro-porous carbon foams with densities ranging from $0.14-0.39$ g cm$^{-3}$ by filter-pressing a mixture of sucrose solutions and particles of sawdust~\cite{2020_chithra_carbonFoam}, newspaper~\cite{Chithra2018carbon-foam-newspaper}, or banana leaf~\cite{Raji2023carbon-foam-banana}, followed by carbonization at high temperature. These porous samples resemble our NP-C and are characterized by turbostratic features with density-independent properties.

As shown in Fig.~\ref{fig:k-rho}(a), the experimentally measured $\kappa$ curves of graphitic carbon foams~\cite{2020_chithra_carbonFoam,Chithra2018carbon-foam-newspaper,Raji2023carbon-foam-banana} all demonstrate a clear linear correlation with density. The similar characteristics of graphene-like fragments and the density-independent behavior between the experimental carbon foams and our NP-C samples result in the same linear $\kappa\propto \rho$ correlation. However, the obvious difference in magnitude between the two can be ascribed to the heterogeneous versus homogeneous geometry of carbon foams and NP-C, respectively. Carbon foams possess mesoscale porous features, originating from the structure of the precursor material, in addition to the nanoscale geometry (i.e., graphitic plane stacking, coordination, defects and curvature). Homogeneous NP-C samples do not possess mesoscale features, but a unimodal distribution of pore sizes which is inextricably linked to, e.g., layer stacking and mid-range order. This is evident when comparing the size of mesoscale pores in carbon foams, of the order of $15$ $\mu$m, \textit{vs.} $5$ nm for NP-C at the same density of $0.3$ g cm$^{-3}$. At the same density, smaller pore sizes imply more connections of graphene sheets in the 3D disordered networks, providing more propagating channels for in-plane heat carriers. This is likely the main reason for the larger $\kappa$ of NP-C as compared to that of carbon foam. This phenomenological relationship between $\kappa$ and $\rho$ is further discussed in Sec.~\ref{Sec:shc} in terms of heat capacity, group velocities and mean free paths. Therefore, while experimental carbon foam data and simulation NP-C data are not amenable to direct comparison, the linear $\kappa (\rho)$ behavior observed in both is indicative of the role of the material nanostructure on the heat conduction characteristics, as we further discuss later in this paper.

For the thermal conductivities of a-C in Fig.~\ref{fig:k-rho}(b), we also included some representative experimental data for comparison~\cite{2000_bullen_a-CKappa,2006_Shamsa_a-CKappa,2021_scott_irradiatteda-C}. Similarly, the $\kappa$ values after quantum-statistical correction are more comparable to the experimental results. Moreover, we fitted the quantum-corrected $\kappa^{\text{a-C}}(\rho)$ curve using high-order polynomials and observe a quadratic dependence. Notably, these experimental density-dependent $\kappa$ values from Bullen \textit{et al.}~\cite{2000_bullen_a-CKappa} and Shamsa \textit{et al.}~\cite{2006_Shamsa_a-CKappa} jointly show the same superlinear trend in Fig.~\ref{fig:k-rho}(b). However, the quantum-corrected $\kappa$ remains slightly higher than their experimental measurements, which is associated with structural differences between computationally generated and experimentally prepared a-C samples. Impurities such as hydrogen or oxygen atoms from chemical vapor deposition and defects like vacancies or voids from physical vapor deposition or sputtering methods can exacerbate scattering and reduce the lifetimes of vibrons. These experimental factors negatively affect the thermal transport properties of a-C. Additionally, as discussed in Sec.~\ref{sec:structure}, our a-C samples prepared by the rapid melt-quench method exhibit slightly higher ordering characteristics in short- and medium-range structures. High ordering enhances the diffusivity of heat carriers and can contribute to higher $\kappa$ in the thermal transport of disordered carbon, which will be discussed in detail below.

In summary, our calculated quantum-corrected thermal conductivities of NP-C and a-C samples generally show qualitative agreement with experimental measurements, particularly in terms of the $\kappa(\rho)$ function\textemdash{}linear correlation for NP-C and a superlinear (quadratic) trend for a-C, thus providing a route to develop a mechanistic understanding of the origin of this difference. In the following sections, we will correlate the physical quantities affecting thermal conductivity with structural characteristics based on the structure-property relationship and ultimately aim to uncover the possible reasons behind the density-dependent thermal conductivities for NP-C and a-C.

\subsubsection{Spectral thermal conductivity}\label{Sec:shc}

\begin{figure*}
    \centering
    \includegraphics[width=1.8\columnwidth]{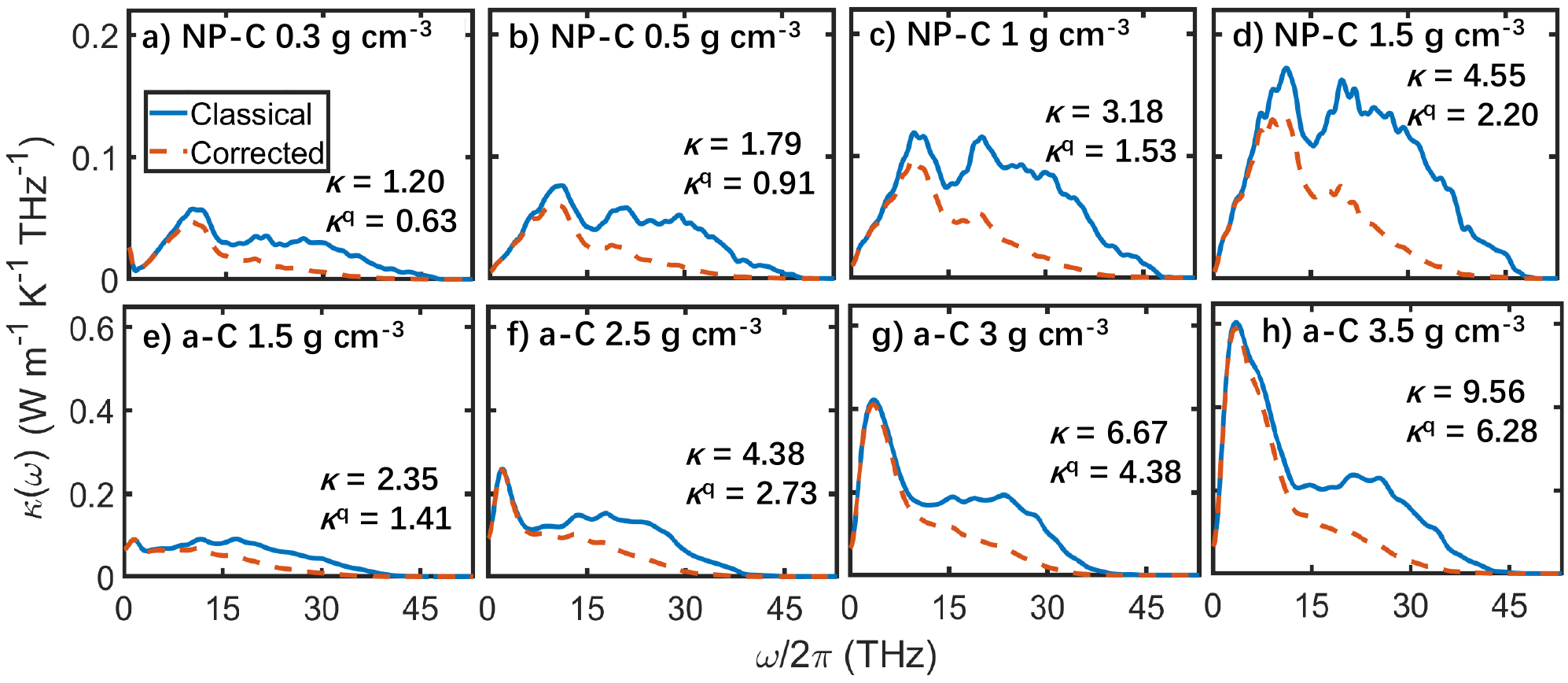}
    \caption{Classical and quantum-corrected spectral thermal conductivity $\kappa(\omega)$  for NP-C with the density of (a) $0.3$, (b) $0.5$, (c) $1$ and (d) $1.5$ g cm$^{-3}$, and for a-C with the density of (e) $1.5$, (f) $2.5$, (g) $3$ and (h) $3.5$ g cm$^{-3}$. Their classical $\kappa$ and quantum-corrected $\kappa^q$ as integral of $\kappa(\omega)$ over vibrational frequency $\omega/2\pi$ are presented in each panel.}
    \label{fig:kw}
\end{figure*}

We first decomposed the total $\kappa$ into frequency-resolved spectral thermal conductivity $\kappa(\omega)$ in \gls{hnemd} calculations at $300$ K. Figure~\ref{fig:kw} presents the classical $\kappa(\omega)$ and quantum-corrected $\kappa^q(\omega)$ for NP-C samples with densities of $0.3$, $0.5$, $1$, and $1.5$ g cm$^{-3}$, and for a-C samples with densities of $1.5$, $2.5$, $3$, and $3.5$ g cm$^{-3}$. Based on Eq.~\eqref{equation:qc_k_omega}, the importance of the quantum correction becomes more pronounced with increasing frequency and reducing temperature. Interestingly, for the classical $\kappa(\omega)$ of NP-C in Fig.~\ref{fig:kw}(a-d), although $\kappa(\omega)$ increases with density, its distributions exhibit a great similarity in profile across different densities, featuring two peaks at $\omega_1/2\pi\approx 10$ and $\omega_2/2\pi\approx 20$ THz. This similarity in profile is independent of the quantum correction, which only proportionally reduces the second peak at high frequency. Additionally, the similarity of the $\kappa(\omega)$ profile can be quantitatively inspected by the approximately constant ratio of $\kappa^q$ to $\kappa$ regardless of density. As seen in Fig.~\ref{fig:kw}(a-d), the ratio holds at approximately $\kappa^q/\kappa\approx0.5$ for NP-C samples, with $\kappa^q/\kappa$ values of $0.53$, $0.51$, $0.48$, and $0.49$ for densities of $0.3$, $0.5$, $1$, and $1.5$ g cm$^{-3}$, respectively.

However, a-C exhibits different features in $\kappa(\omega)$ as compared to NP-C. As shown in Fig.~\ref{fig:kw}(e-h), unlike the proportional scaling of the $\kappa$ ($\kappa^q$) profile for NP-C, the first peak of $\kappa(\omega)$ of a-C at low frequency ($\omega_1/2\pi \approx 3$ THz) scales up disproportionately with density, showing a biased growth relative to the second peak at $\omega_2/2\pi \approx 20$ THz. This implies a weaker effect of quantum correction on $\kappa$ of a-C as compared to NP-C. Accordingly, as seen in Fig.~\ref{fig:kw}(e-h), a-C has a larger $\kappa^q/\kappa$ value compared to NP-C, with increasing ratios of $0.6$, $0.62$, $0.65$, and $0.66$ for densities of $1.5$, $2.5$, $3$, and $3.5$ g cm$^{-3}$, respectively. This indicates that the $\kappa(\omega)$ distribution of a-C is influenced by its density, with higher densities exhibiting enhanced vibrational modes at lower frequencies.

The similarity in the $\kappa$ profile for NP-C likely originates from its density-independent structural characteristics. According to the phonon gas model, lattice thermal conductivity scales as $\kappa \propto C \rho v_{\rm g} \lambda$, where $C$ is the specific heat capacity per unit mass (the product of $C \rho$ denotes the volumetric heat capacity $C_V$), $v_{\rm g}$ is the group velocity, and $\lambda$ is the mean free path (MFP). $v_{\rm g}$ is typically determined by the stiffness of chemical bonds, while $\lambda$ is related to the ordering of atomic arrangement. The latter two quantities depend on the structural characteristics. The insensitivity of structural characteristics, such as the fraction of sp$^2$ bonds and short- and medium-range orderings, to density, as well as the proportional scaling of $C_V$ with density, likely explain the linear $\kappa \propto \rho$ relationship for NP-C.

However, the underlying factors dictating the superlinear $\kappa(\rho)$ relationship for a-C appear to be more complex. In addition to the contribution of $C_V$ from mass density in the phonon gas model, the increasing $\kappa^q/\kappa$ ratio with density suggests that certain structural characteristics have enhanced the diffusivity ($\kappa/C_V$ or $v_{\rm g}\lambda$) of vibrons in thermal transport. Our hypothesis is that the sp$^3$ motifs may be the key factor in increasing the $\kappa (\omega)$ at low frequencies. In fact, it has been computationally shown that sp$^3$ motifs act as propagons (phonon-like vibrons) to strengthen the diffusivity of low-frequency vibrational modes~\cite{2022_npjcm_giri_CN-kappa,moon2024crystal}. Based on the discussion of correlations between structural properties and thermal conductivity above,  we hypothesize that both $v_{\rm g}$ and $\lambda$ are independent of density for NP-C, while the diffusivity ($v_{\rm g}\lambda$) increases with density for a-C. We will assess the validity of these hypotheses in detail in the following sections.

\subsubsection{Mean free paths}

\begin{figure}
    \centering
    \includegraphics[width=1\columnwidth]{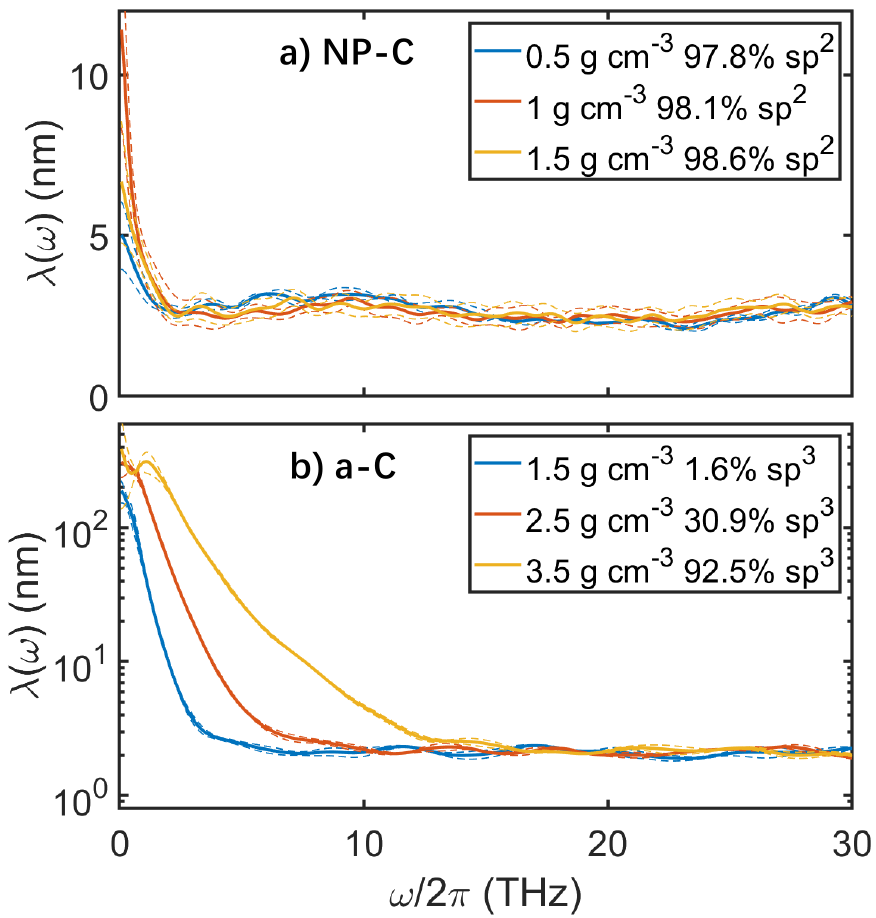}
    \caption{Vibrational mean-free paths $\lambda(\omega)$ for (a) NP-C (of sp$^2$ fraction virtually independent of density) with densities of $0.5$, $1$ and $1.5$ g cm$^{-3}$, and (b) a-C (of sp$^3$ content varying with density) with densities of $1.5$, $2.5$ and $3.5$ g cm$^{-3}$. Three independent runs are averaged to obtain error bounds (adjoint thin lines) for each density.}
    \label{fig:mfp}
\end{figure}

\begin{figure}
    \centering
    \includegraphics[width=1\columnwidth]{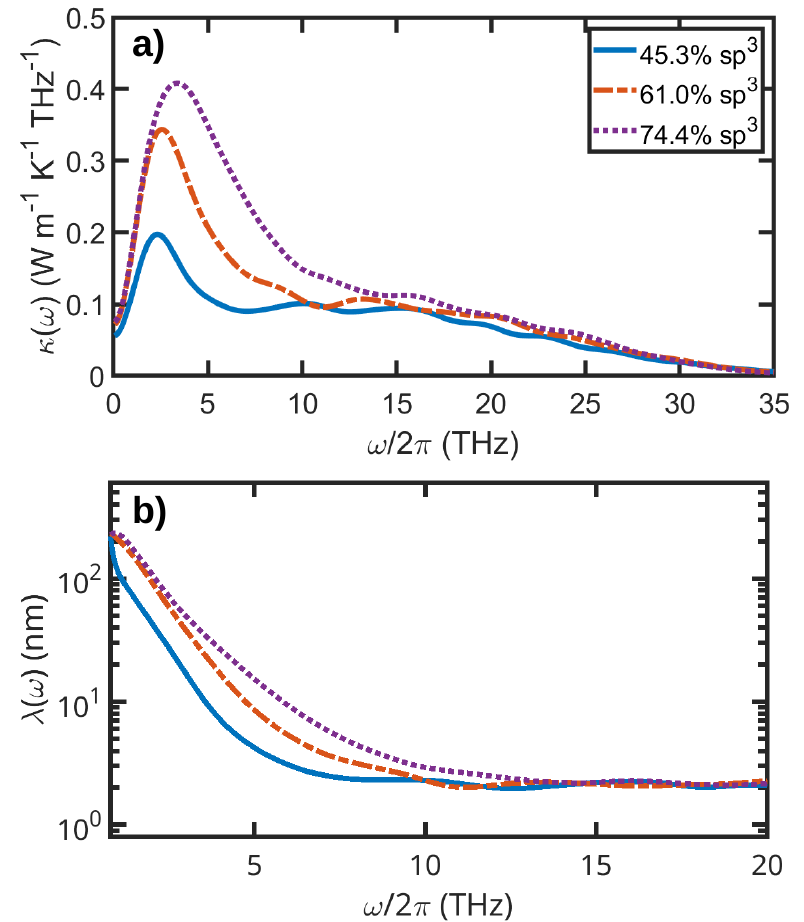}
    \caption{(a) Quantum-corrected spectral thermal conductivity $\kappa(\omega)$ and (b) vibrational mean-free path $\lambda(\omega)$ versus sp$^3$ content in a-C with a fixed density of $3$ g cm$^{-3}$. Three independent runs are conducted for average for each case.}
    \label{fig:rho3}
\end{figure}

In order to investigate the role of MFPs in thermal transport, we performed \gls{nemd} calculations for NP-C with densities of $0.5$, $1$, and $1.5$ g cm$^{-3}$, and a-C with densities of $1.5$, $2.5$, and $3.5$ g cm$^{-3}$. Their ballistic thermal conductances $G(\omega)$ at low temperatures are plotted in Fig. S5 and S6. Using Eq.~\eqref{eq:mfp} we obtained the frequency-resolved $\lambda(\omega)$ for NP-C and a-C. As seen in Fig. \ref{fig:mfp}(a), NP-C shows insensitivity of the MFP to density due to similar structural characteristics. In contrast, the a-C results reveal a clear dependence of the MFPs in the low-frequency domain in Fig. \ref{fig:mfp}(b), with higher density enlarging the domain of $\lambda(\omega)$. Specifically, using an MFP of $3$ nm as a benchmark, its frequency threshold extends roughly from $4$ THz at $1.5$ g cm$^{-3}$ to $13$ THz at $3.5$ g cm$^{-3}$. Note that, as discussed above, density itself merely contributes to the heat capacity $C_V$. We thus speculate that the enhancement of $\lambda(\omega)$ is caused by the increased fraction of sp$^3$ in a-C.

To validate this assumption, it is necessary to isolate the sp$^3$ content from density. To achieve this, we generated three types of a-C samples with different sp$^3$ fractions at a fixed density using the following strategy. Due to the dependence of sp$^3$ on density, we first generated samples with densities of $2.75$, $3$, and $3.25$ g cm$^{-3}$, each corresponding to specific sp$^3$ fractions. Next, we performed deformed \gls{md} simulations under room temperature with a time step of 1 fs. Specifically, we uniformly deformed the boxes of the highest and lowest densities to adjust their volumes to the same $3$ g cm$^{-3}$. The $3.25$ g cm$^{-3}$ box was slowly stretched at a deformation rate of $10^{-5}$ \AA~fs$^{-1}$ for $0.2475$ ns, while a similar compression operation was applied to the $2.75$ g cm$^{-3}$ a-C samples for $0.2766$ ns. Finally, we reequilibrated and relaxed these samples at $300$ K for $1$ ns to remove metastabilities.

We performed similar \gls{hnemd} (see Fig. S7 and S8) and \gls{nemd} (see Fig. S9) calculations on these samples and obtained their quantum-corrected spectral thermal conductivities and MFPs. As observed in Fig.~\ref{fig:rho3}(a), the $\kappa(\omega)$ peak swells significantly around $3$ THz, increasing from $0.2$ to $0.4$ W m$^{-1}$ K$^{-1}$ THz$^{-1}$, with the total quantum-corrected $\kappa^q$ rising from $2.56$ to $4.41$ W m$^{-1}$ K$^{-1}$ as the sp$^3$ fraction increases from $45.3\%$ to $74.4\%$. This indicates that, apart from the volumetric heat capacity linearly increasing $\kappa(\omega)$ across the entire frequency domain, the sp$^3$ motifs preferentially strengthen $\kappa(\omega)$ at low frequencies. We further present the effect of sp$^3$ on $\lambda(\omega)$ at a fixed $3$ g cm$^{-3}$ in Fig.~\ref{fig:rho3}(b). As discussed in Fig.~\ref{fig:mfp}(b), sp$^3$ motifs extend the $\kappa(\omega)$ distribution to a larger domain at low frequencies. At $\lambda=3$ nm, the frequency threshold ($\omega/2\pi$) extends from $5$ to $10$ THz as the sp$^3$ fraction increases from $45.3\%$ to $74.4\%$. Therefore, from Fig.~\ref{fig:rho3} we clearly conclude that sp$^3$ motifs enhance low-frequency $\lambda(\omega)$, which explains the disproportionate rise of the $\kappa(\omega)$ peak in the low-frequency domain in Fig.~\ref{fig:kw}(e-h).

\subsubsection{Group velocity}

The correlations of (spectral) thermal conductivity and MFP with structural characteristics have been discussed above. In this section, we continue to study the group velocities from phonon dispersions for the NP-C and a-C structures with different densities. Because of the lack of a well-defined wave vector due to the non-periodicity in disordered materials, it is in principle impossible to obtain a well-defined phonon dispersion. However, we can still investigate the dispersion of low-frequency vibrational modes, as these acoustic modes usually exhibit phonon-like behavior such as a well-defined group velocity $v_{\rm g}$.

We fitted the acoustic $v_{\rm g}$ values from current correlation function, which is considered as the spatially dependent generalization of velocity correlations \cite{2021_dynasor}. The definition of this function starts with the atom density
\begin{equation}
    n(\mathbf{r},t)=\sum_i^N \delta[\mathbf{r} - \mathbf{r}_i(t)],
\end{equation}
where $\mathbf{r}_i(t)$ is the position of atom $i$ at time $t$, and $N$ is the number of system atoms. The current density $\mathbf{j}(\mathbf{r},t)$ is further expressed as the product of atom velocity $\mathbf{v}_i(t)$ and atom density $n(\mathbf{r},t)$:
\begin{equation}
    \mathbf{j}(\mathbf{r},t) = \sum_i^N \mathbf{v}_i(t) \delta[\mathbf{r} - \mathbf{r}_i(t)],
\end{equation}
which can be spatially Fourier transformed as
\begin{equation}
    \mathbf{j}(\mathbf{q},t) = \sum_i^N \mathbf{v}_i(t) e^{\text{i}\mathbf{q} \cdot \mathbf{r}_i(t)}.
\end{equation}
It can be decomposed into longitudinal and transverse components, which are parallel and perpendicular to the $\mathbf{q}$ vector, respectively:
\begin{equation}
    \mathbf{j}(\mathbf{q},t) = \mathbf{j}_{\rm L}(\mathbf{q},t) + \mathbf{j}_{\rm T}(\mathbf{q},t),
\end{equation}
where
\begin{equation}
    \begin{split}
          \mathbf{j}_{\rm L}(\mathbf{q},t) = \sum_i^N [\mathbf{v}_i(t) \cdot \hat{\mathbf{q}}] \hat{\mathbf{q}} e^{\text{i}\mathbf{q} \cdot \mathbf{r}_i(t)};  \\  
        \mathbf{j}_{\rm T}(\mathbf{q},t) = \sum_i^N \{\mathbf{v}_i(t) - [\mathbf{v}_i(t) \cdot \hat{\mathbf{q}}] \hat{\mathbf{q}} \} e^{\text{i}\mathbf{q} \cdot \mathbf{r}_i(t)}.
    \end{split}
\end{equation}
The corresponding current correlation functions can now be computed as
\begin{equation}
    \begin{split}
        C_{\rm L}(\mathbf{q},t) = \frac{1}{N} \langle \mathbf{j}_{\rm L}(\mathbf{q},t) \cdot \mathbf{j}_{\rm L}(-\mathbf{q},0) \rangle; \\
        C_{\rm T}(\mathbf{q},t) = \frac{1}{N} \langle \mathbf{j}_{\rm T}(\mathbf{q},t) \cdot \mathbf{j}_{\rm T}(-\mathbf{q},0) \rangle,
    \end{split}
\end{equation}
and their temporal Fourier transforms are expressed as
\begin{equation}
    \begin{split}
        C_{\rm L}(\mathbf{q},\omega) = \int_{-\infty}^{\infty} C_{\rm L}(\mathbf{q},t)e^{-\text{i}\omega t} \text{d}t; \\
        C_{\rm T}(\mathbf{q},\omega) = \int_{-\infty}^{\infty} C_{\rm T}(\mathbf{q},t)e^{-\text{i}\omega t} \text{d}t.
    \end{split}
\end{equation}

Figures S10-S12 and S13-S15 compare the longitudinal and transverse current correlation functions in the ($\omega$, $|\mathbf{q}|$) plane for NP-C and a-C samples. We extracted their group velocities in Table \ref{tab:vg} by fitting the ratio of $\omega$ to $|\mathbf{q}|$ at low frequencies. For the NP-C samples, as discussed in Sec.~\ref{Sec:shc}, $v_{\rm g}$ is primarily governed by the nature of sp$^2$ bonds, and we see that density-insensitive sp$^2$ ratio causes longitudinal and transverse $v_{\rm g}$ components to remain around $1.7$ and $0.7$ km s$^{-1}$, respectively. However, a-C demonstrates a sensitivity of group velocity to its density due to the change in sp$^3$ content. The $91\%$ increase of sp$^3$ fraction from $1.5$ to $3.5$ g cm$^{-3}$ enhances group velocity with an $1.5$-fold ($2$-fold) increase in the longitudinal (transverse) component. Finally, the higher group velocity of a-C compared to NP-C indicates a positive effect of sp$^3$ motif on thermal conductivity.

\begin{figure}
    \centering
    \includegraphics[width=1\columnwidth]{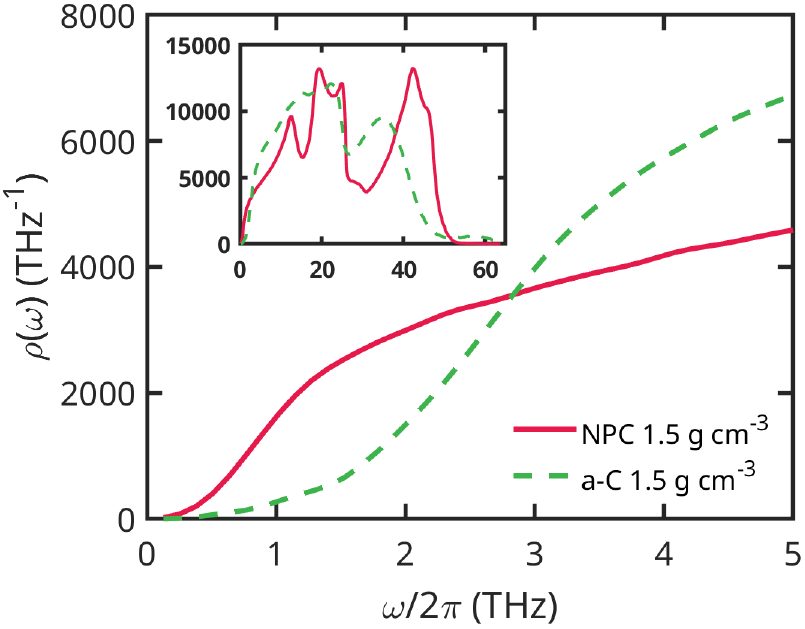}
    \caption{Vibrational density of states $\rho(\omega)$ for NP-C and a-C at the same density of $1.5$~g~cm$^{-3}$ in the low-frequency range ($\omega/2\pi = 1$–$5$~THz). The inset shows the total VDOS distribution over the full frequency range.}
    \label{fig:VDOS}
\end{figure}

It is worth noting that at the same density of $1.5$~g~cm$^{-3}$ a-C exhibits a larger MFPs than NP-C in the low-frequency domain. However, a-C has a lower thermal conductivity ($1.4$~W~m$^{-1}$~K$^{-1}$ for a-C \textit{vs.} $2.2$~W~m$^{-1}$~K$^{-1}$ for NP-C). This is attributed to the significantly lower vibrational density of states (VDOS) for a-C as compared to NP-C, which leads to a smaller heat capacity (integral of VDOS) in the low-frequency region. The vibrational density of states, $\rho(\omega)$, was calculated using the mass-weighted velocity autocorrelation function $C_m(t)$ through Fourier transformation:
\begin{equation}
    \rho(\omega) = \int_{-\infty}^{+\infty} C_m(t) e^{i\omega t} \text{d}t,
\end{equation}
where
\begin{equation}
    C_m(t) = \frac{1}{k_\text{B} T} \sum_{i=1}^{N} m_i \langle \vec{v}_i(t) \cdot \vec{v}_i(0) \rangle,
\end{equation}
where $k_\text{B}$ and $T$ are the Boltzmann constant and temperature, respectively, and $m_i$ and $\vec{v}_i$ represent the mass and velocity of particle $i$ in the $N$-atom system. Figure~\ref{fig:VDOS} shows the calculated VDOS for both a-C and NP-C at the same density of $1.5$~g~cm$^{-3}$. In the low-frequency region ($\omega/2\pi < 3$~THz), NP-C exhibits a significantly larger VDOS than in a-C. Furthermore, as shown in Fig.~\ref{fig:kw}(e), the excess quantum-corrected thermal conductivity of NP-C in the low-frequency region, relative to a-C in Fig.~\ref{fig:kw}(d), aligns with its higher VDOS in this frequency range. This consistency between the VDOS and the frequency-resolved thermal conductivity confirms that the smaller VDOS at low frequencies is the primary reason for the lower thermal conductivity of a-C as compared to NP-C.

\begin{table}[]
    \centering
    \caption{Fitted longitudinal and transverse group velocity components from the calculations of corresponding current correlation functions for the NP-C and a-C samples with different densities (g cm$^{-3}$). Statistical averages and standard deviations (shown in parentheses) were obtained from three independent runs for each density. The units of $v_{\rm g}$ are km s$^{-1}$.}
    \begin{tabular}{p{1cm} p{1cm} p{1cm} p{1cm} p{1cm} p{1cm} p{1cm}}
    \hline
    \hline
         & \multicolumn{3}{c}{Longitudinal} & \multicolumn{3}{c}{Transverse}  \\
    \hline
         NP-C & 0.5 & 1 & 1.5 & 0.5 & 1 & 1.5 \\
         $v_{\rm g}$ & 1.76 (0.06) & 1.57 (0.06) & 1.75 (0.11)  & 1.09 (0.03) & 0.93 (0.08) & 0.96 (0.06) \\
    \hline
         a-C & 1.5 & 2.5 & 3.5 & 1.5 & 2.5 & 3.5 \\
         $v_{\rm g}$ & 1.31 (0.08) & 1.89 (0.09) & 2.91 (0.07)  & 1.08 (0.06) & 1.22 (0.02) & 1.27 (0.07) \\
    \hline
    \hline
    \end{tabular}
    \label{tab:vg}
\end{table}

\subsection{Effect of structural ordering}\label{sec:order}

\begin{figure}
    \centering
    \includegraphics[width=1\columnwidth]{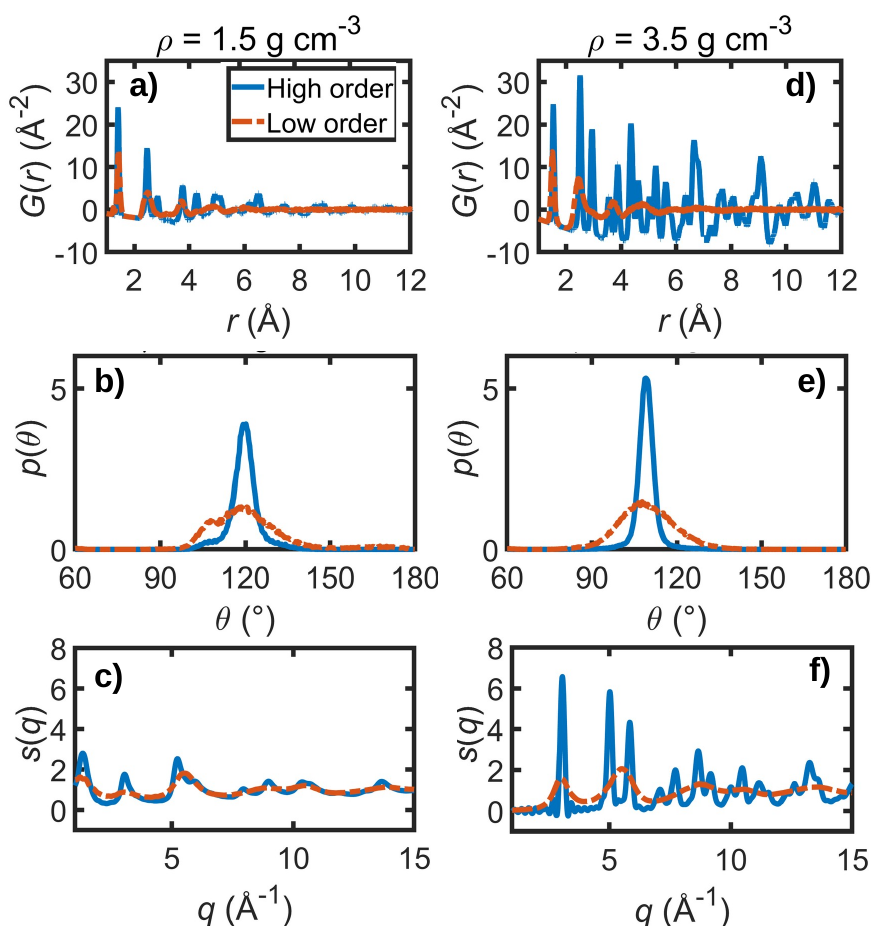}
    \caption{(a, d) Reduced pair correlation functions $G(r)$, (b, e) angular distribution functions $p(\theta)$ and (c, f) structure factors $s(q)$ for the low- and high-ordered 8000-atom a-C structures with densities of $1.5$ (a-c) and $3.5$ (d-f) g cm$^{-3}$.}
    \label{fig:stru-order}
\end{figure}

Structural ordering has a significant impact on thermal conductivity. In this section, we selected the two smallest and largest densities of 8000-atom a-C models to investigate this effect. Choosing quenching rates of $10^{12}$ and $10^{15}$ K s$^{-1}$ between $5000$ and $1000$ K in the temperature protocol of Fig.~\ref{fig:temperature}, we prepared a-C samples with different degrees of structural order. Figure~\ref{fig:stru-order} compares the structural properties of those samples by analyzing their ADFs, structure factors and reduced PCFs, $G(r) = 4 \pi n r[g(r)-1]$. We observe that a-C samples prepared at a low-quenching rate all exhibit more (in larger $r$ or $q$) and sharper peaks in $G(r)$, $p(\theta)$, and $s(q)$, compared to those prepared at a high-quenching rate. These results indicate that the carbon structures prepared at a low-quenching rate in Fig.~\ref{fig:stru-order} have more ordered characteristics in both the short and medium range.

\begin{figure}
    \centering
    \includegraphics[width=1\columnwidth]{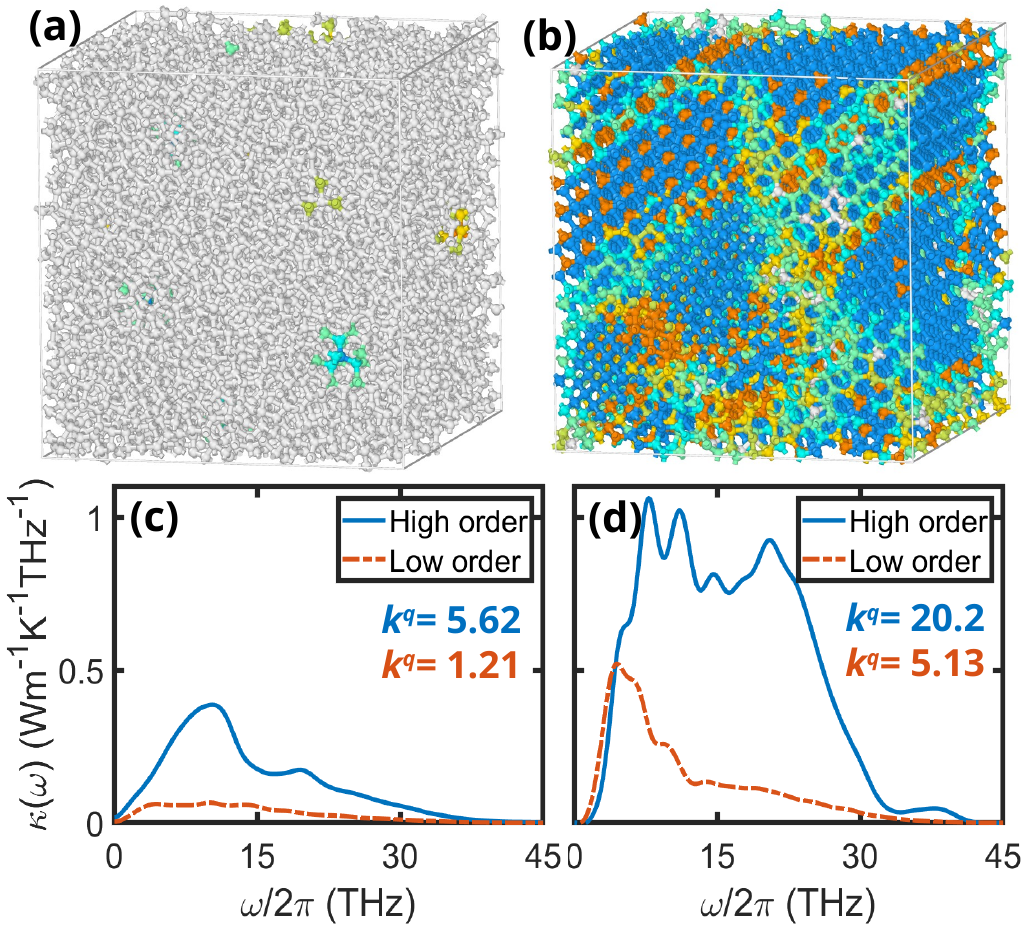}
    \caption{Identification of diamond configuration for (a) Low-ordering and (b) high-ordering $3.5$ g cm$^{-3}$ 8000-atom a-C snapshots. Cubic diamond, hexagonal diamond pockets and others are identified by extended Common Neighbor Analysis method~\cite{extended-CNA_2016} in blue, orange and gray, respectively. Quantum corrected spectral thermal conductivity of high-ordering and low-ordering a-C samples of (c) $1.5$ and (d) $3.5$ g cm$^{-3}$. Twelve independent runs were performed to average $\kappa(\omega)$ for each density.}
    \label{fig:kappa-order}
\end{figure}

We inspected $3.5$ g cm$^{-3}$ carbon samples in detail and analyzed the degree of crystallization (diamond-like local motifs) for high-order and low-order samples. In Fig.~\ref{fig:kappa-order}(a-b), cubic diamond, hexagonal diamond, and other motifs are rendered in blue, orange, and gray, respectively. Compared with the low-order sample in Fig.~\ref{fig:kappa-order}(a), the high-order $3.5$ g cm$^{-3}$ sample is more crystalline and locally forms a few tiny (size of $\approx 0.8$ nm) paracrystalline diamond pockets, featuring a semi-crystalline structure. We further calculated the quantum-corrected $\kappa(\omega)$ for the two densities in Fig.~\ref{fig:kappa-order}(c-d). Structural order has a significant impact on thermal conductivity, and we observe that the high-order sample leads to a $4$- or $5$-fold increase in $\kappa^q$ for both densities compared to the low-order sample. High order (more tiny crystalline pockets) in the semi-crystalline structure increases its MFP~\cite{2021_Shang_nature_a-C} and then activates more heat carriers in $\kappa(\omega)$ across the entire frequency domain, leading to a significant increase in thermal conductivity. Experimental measurements indicate that the formation of diamond-like medium-range order (MRO) clusters in the amorphous matrix of ultra-hard a-C samples, prepared by heating compressed fullerene, significantly enhances thermal conductivity~\cite{2021_Shang_nature_a-C}. Notably, the thermal conductivity can be as high as $\kappa = 20-30$~W~m$^{-1}$~K$^{-1}$ when the MRO clusters grow to a few nanometers in size. This observation aligns with our conclusion that structural ordering plays a critical role in improving thermal transport.

\section{Summary and Conclusions}

In this paper, we have systematically investigated the structural properties and thermal transport of disordered carbon over a wide range of densities using molecular dynamics simulations with neural evolution-based machine learning interatomic potentials. Utilizing the melt-graphitization-quench and rapid melt-quench \gls{md} protocols, we prepared $125000$-atom (size of $10-20$ nm) NP-C samples with densities varying from $0.3$ to $1.5$ g cm$^{-3}$ and a-C samples with densities from $1.5$ to $3.5$ g cm$^{-3}$. The structural features of these disordered samples, comparable to experimental ones, were carefully characterized through analyses of morphologies, atomic coordinations, PCFs, ADFs, and structure factors. Furthermore, combining the calculations of \gls{hnemd}, \gls{nemd}, and current correlation function, we comprehensively studied the thermal transport properties of NP-C and a-C structures. We correlated fundamental quantities associated with thermal transport, such as frequency-resolved thermal conductivity, MFPs, and dispersion relations, with their structural characteristics.

Specifically, NP-C structures with a predominant $98\%$ fraction of sp$^2$ bonds consist of curved graphene fragments assembled into 3D networks. Their structural characteristics in short and medium ranges are similar and show insensitivity to their densities. Due to the insensitivity of group velocity and MFPs, as well as the heat capacity $C_V$ being proportional to mass density, our calculated quantum-corrected thermal conductivity $\kappa$ shows a linear dependence on density. This agrees well with experimental measurements of $\kappa$ for carbon foam.

Moreover, the a-C structure shows a positive correlation of sp$^3$ content with density, increasing from $2\%$ at $1.5$ g cm$^{-3}$ to $93\%$ at $3.5$ g cm$^{-3}$. The short- and medium-range ordering characteristics of a-C structures were analyzed and compared well (though slightly more ordered) with experimental results~\cite{1995_Gilkes_prb_aC}. Thermal conductivity \textit{vs} density scales as a superlinear $\kappa(\rho)$ relation, consistent with experimental measurements~\cite{2000_bullen_a-CKappa,2006_Shamsa_a-CKappa}. We attribute the superlinear rise in $\kappa$ to the sp$^3$ content, which enhances $\kappa(\omega)$ from low-frequency vibrational modes. Finally, we also show that structural order significantly enhances $\kappa$.

\vspace{0.5cm}
\noindent{\textbf{Data availability}}

The inputs and outputs related to the \gls{nep} model training are freely available at the Gitlab repository: \url{https://gitlab.com/brucefan1983/nep-data}. The datasets, inputs and outputs of \gls{md} simulations, as well as the relevant scripts for processing and plotting, are freely available at Zenodo \cite{data_zenodo}.

\vspace{0.5cm}
\acknowledgments

The authors acknowledge funding from the Academy of Finland,
under projects 312298/QTF
Center of Excellence program (T.A-N., \& Y.W.), 321713 (M.A.C. \& Y.W.), 330488 (M.A.C.). T.A-N. and Y.W. have also been supported in part by Academy of Finland’s grant no. 353298 under the European Union -- NextGenerationEU instrument.The authors acknowledge computational resources from the Finnish Center for Scientific Computing (CSC) and Aalto University's Science IT project. Y.W thanks Mr. Ting Liang for the discussion of current correlation function calculations.

\bibliography{refs}

\end{document}


\maketitle

\begin{figure}[h]
\addcontentsline{toc}{subsection}{\autoref{fig:figs_NEP-train} Trained NEP}
\centering
\includegraphics[width=0.9\columnwidth]{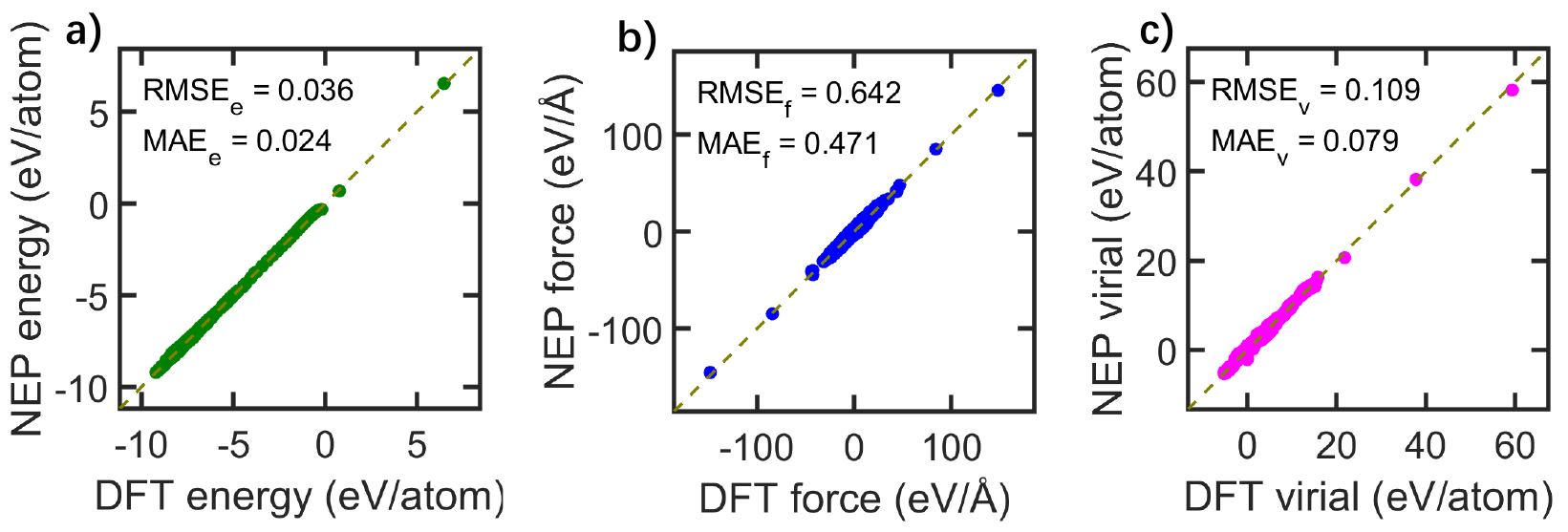}
\caption{Parity plots of (a) energy and (b) force and (c) virial between  NEP predictions and DFT-PBE reference data for the training dataset. Root-mean-square errors (RMSEs) and mean absolute errors (MAEs) are indicated in each subplot.}
\label{fig:figs_NEP-train}
\end{figure}

\begin{figure}
\addcontentsline{toc}{subsection}{\autoref{figs:size-effect} Effect of thermal conductivity on model size}
\centering
\includegraphics[width=0.55\columnwidth]{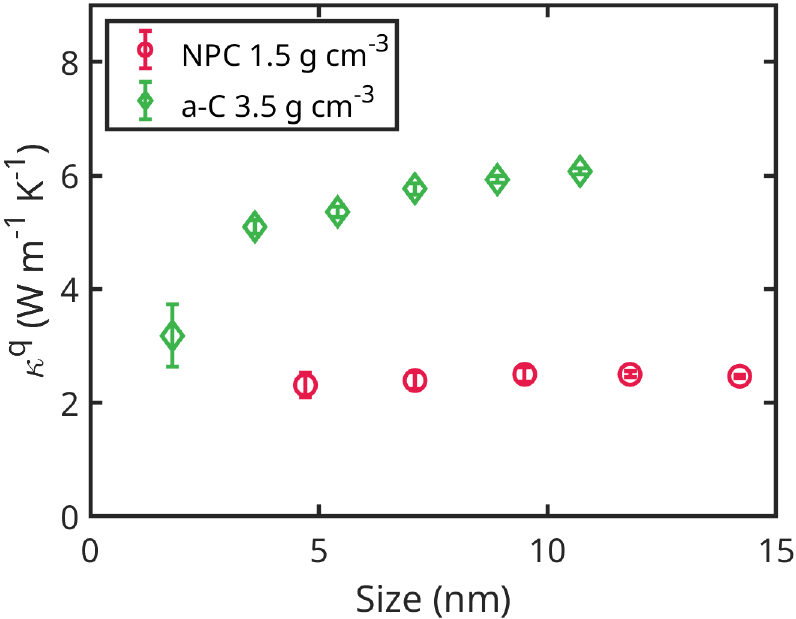}
\caption{Effect of quantum-corrected thermal conductivity ($\kappa^\text{q}$) on model size. The $1.5$~g~cm$^{-3}$ NP-C and $3.5$~g~cm$^{-3}$ a-C samples were selected for the size effect test. For each model size, nine independent simulations were performed to obtain statistical averages.}
\label{figs:size-effect}
\end{figure}

\begin{figure}
\addcontentsline{toc}{subsection}{\autoref{figs:kappa-time-NP-C}  Cumulative $\kappa$ average with HNEMD production time for NP-C with different densities}
\centering
\includegraphics[width=0.9\columnwidth]{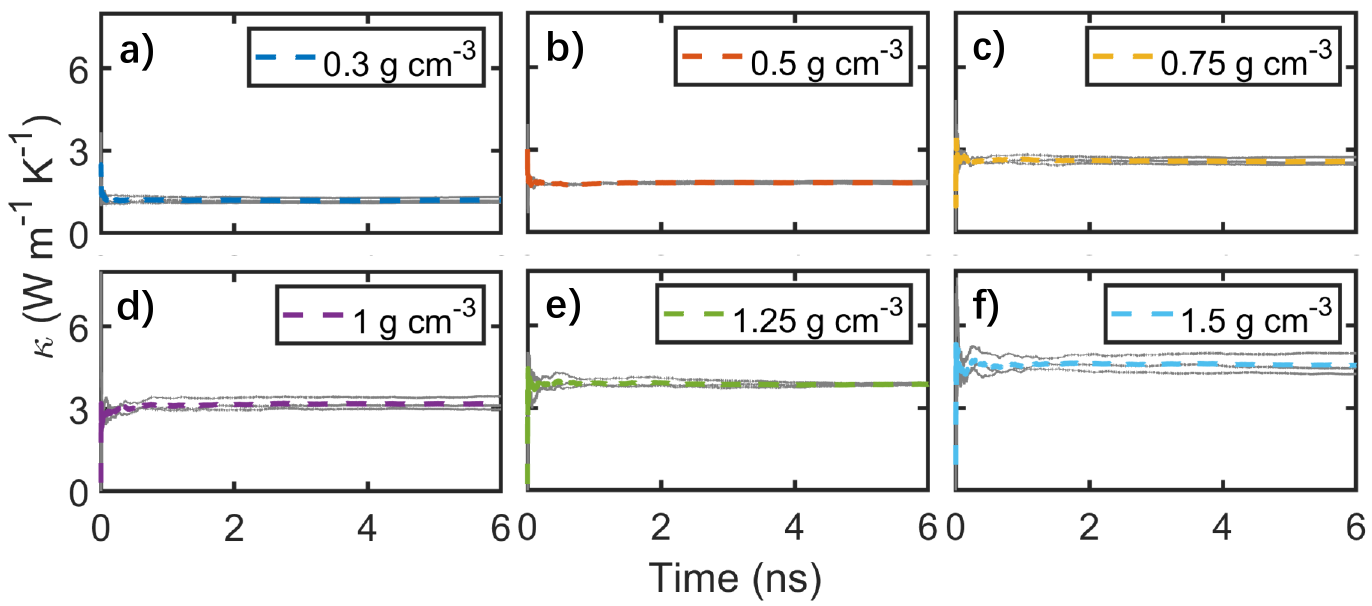}
\caption{
  Cumulative average of thermal conductivity $\kappa$ as a function of the HNEMD production time for NP-C samples with the densities of (a) $0.3$, (b) $0.5$, (c) $0.75$, (d) $1$, (d) $1.25$ and (f) $1.5$ g cm$^{-3}$. The thin solid lines represent results from three independent runs and the thick dashed line in each subplot is the average.}
\label{figs:kappa-time-NP-C}
\end{figure}

\begin{figure}
\addcontentsline{toc}{subsection}{\autoref{figs:kappa-time-a-C} Cumulative $\kappa$ average with HNEMD production time for a-C with different densities}
\centering
\includegraphics[width=1\columnwidth]{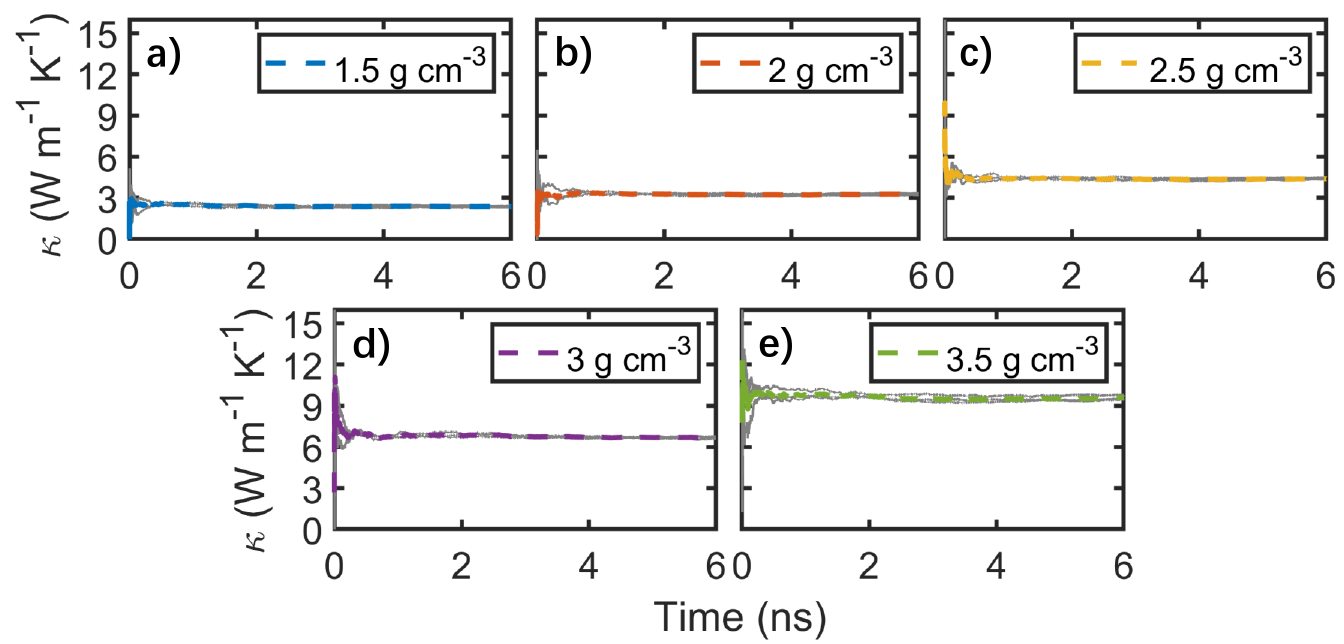}
\caption{
  Cumulative average of thermal conductivity $\kappa$ as a function of the HNEMD production time for a-C samples with the density of (a) $1.5$, (b) $2$, (c) $2.5$, (d) $3$ and (e) $3.5$ g cm$^{-3}$. The thin solid lines represent results from three independent runs and the thick dashed line in each subplot is the average.}
\label{figs:kappa-time-a-C}
\end{figure}

\begin{figure}
\addcontentsline{toc}{subsection}{\autoref{figs:Gw-vs-density_NP-C}  Spectral thermal conductance $G(\omega)$ for NP-C with different densities}
\centering
\includegraphics[width=0.5\columnwidth]{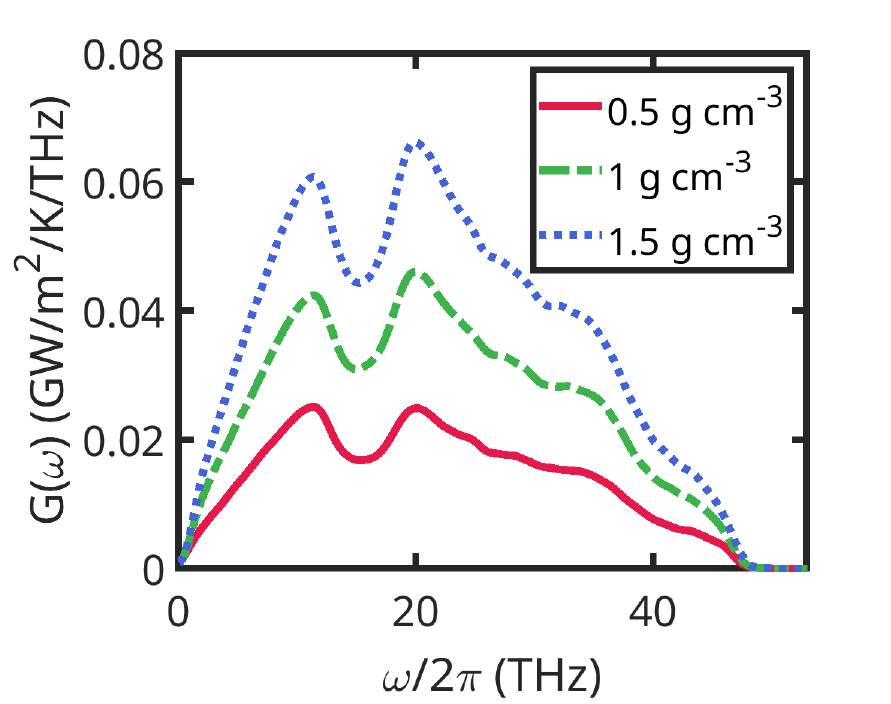}
\caption{
  Classical spectral thermal conductance $G(\omega)$ of NP-C with the different densities. The results are averaged over three independent runs for each density.}
\label{figs:Gw-vs-density_NP-C}
\end{figure}

\begin{figure}
\addcontentsline{toc}{subsection}{\autoref{figs:Gw-vs-density_a-C}  Spectral thermal conductance $G(\omega)$ for a-C with different densities}
\centering
\includegraphics[width=0.5\columnwidth]{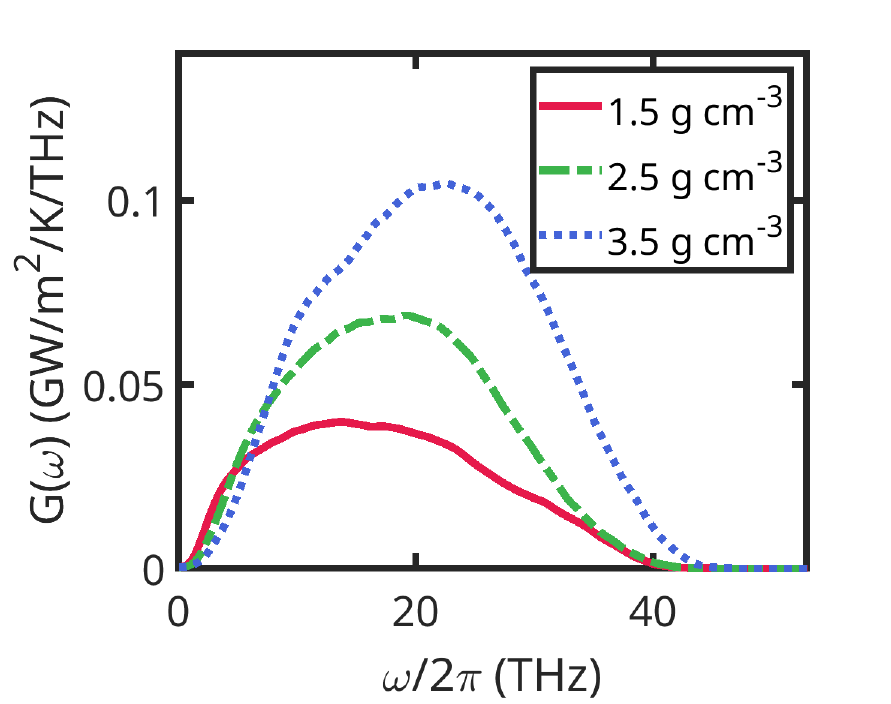}
\caption{
  Classical spectral thermal conductance $G(\omega)$ of a-C with the different densities. The results are averaged over three independent runs for each density.}
\label{figs:Gw-vs-density_a-C}
\end{figure}

\begin{figure}
\addcontentsline{toc}{subsection}{\autoref{figs:rho3-sp3-varing-kappa-vs-time}  Cumulative $\kappa$ average with HNEMD production time for a-C of $3$ g cm$^{-3}$ with different sp$^3$ fractions}
\centering
\includegraphics[width=1\columnwidth]{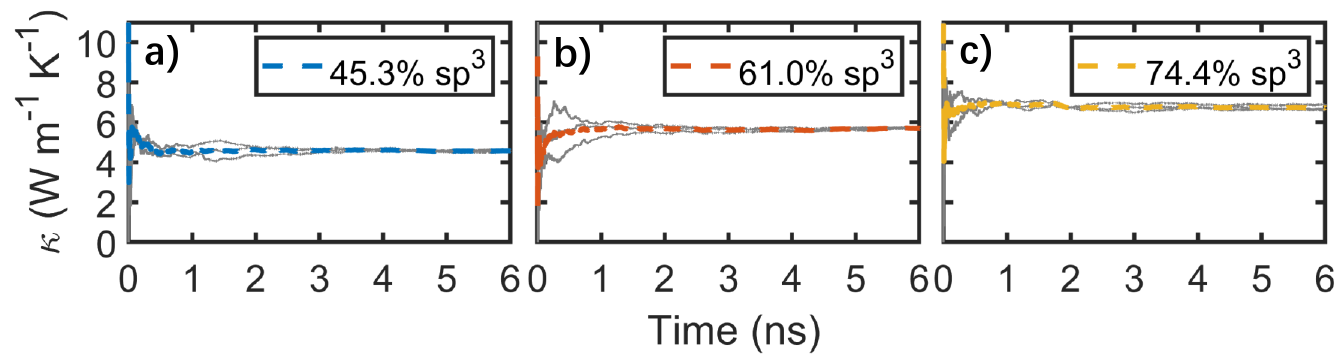}
\caption{
  Cumulative average of the thermal conductivity $\kappa$ as a function of the HNEMD production time for a-C samples with the sp$^3$ fractions of (a) $45.3\%$, (b) $61.0\%$ and (c) $74.4\%$ at a fixed density of $3$ g cm$^{-3}$. The thin solid lines represent results from three independent runs and the thick dashed line in each subplot is the average.}
\label{figs:rho3-sp3-varing-kappa-vs-time}
\end{figure}

\begin{figure}
\addcontentsline{toc}{subsection}{\autoref{figs:averaged_kw_sp3-varing}  Spectral thermal conductivity $\kappa(\omega)$ for a-C of $3$ g cm$^{-3}$ with different sp$^3$ fractions}
\centering
\includegraphics[width=1\columnwidth]{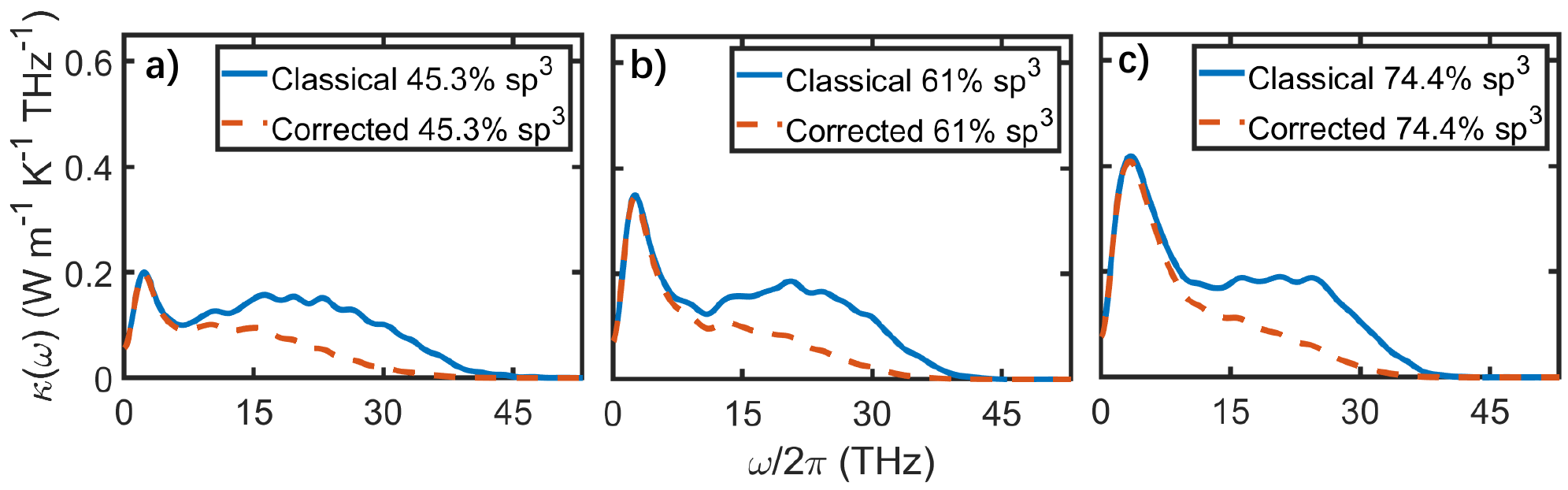}
\caption{
  Classical and quantum-corrected spectral thermal conductivity $\kappa(\omega)$ of a-C with a fixed density of $3$ g cm$^{-3}$ but different sp$^3$ fractions: (a) $45.3\%$, (b) $61.0\%$ and (c) $74.4\%$. The results are averaged over three independent runs for each case.}
\label{figs:averaged_kw_sp3-varing}
\end{figure}

\begin{figure}
\addcontentsline{toc}{subsection}{\autoref{figs:Gw-vs-sp3}  Spectral thermal conductance $G(\omega)$ for a-C of $3$ g cm$^{-3}$ with different sp$^3$ fractions}
\centering
\includegraphics[width=0.5\columnwidth]{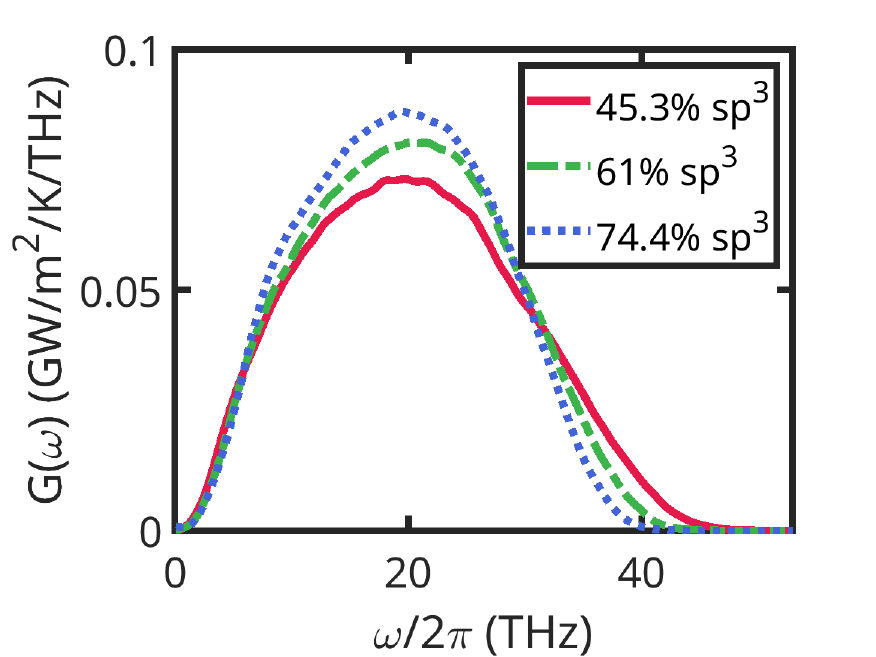}
\caption{
  Classical spectral thermal conductance $G(\omega)$ of a-C with the different sp$^3$ fractions at same density of $3$ g cm$^{-3}$. The results are averaged over three independent runs for each case.}
\label{figs:Gw-vs-sp3}
\end{figure}

\begin{figure}
\addcontentsline{toc}{subsection}{\autoref{figs:NP-C-ccf-0.5}  Current correlation functions for $0.5$ g cm$^{-3}$ NP-C}
\centering
\includegraphics[width=1\columnwidth]{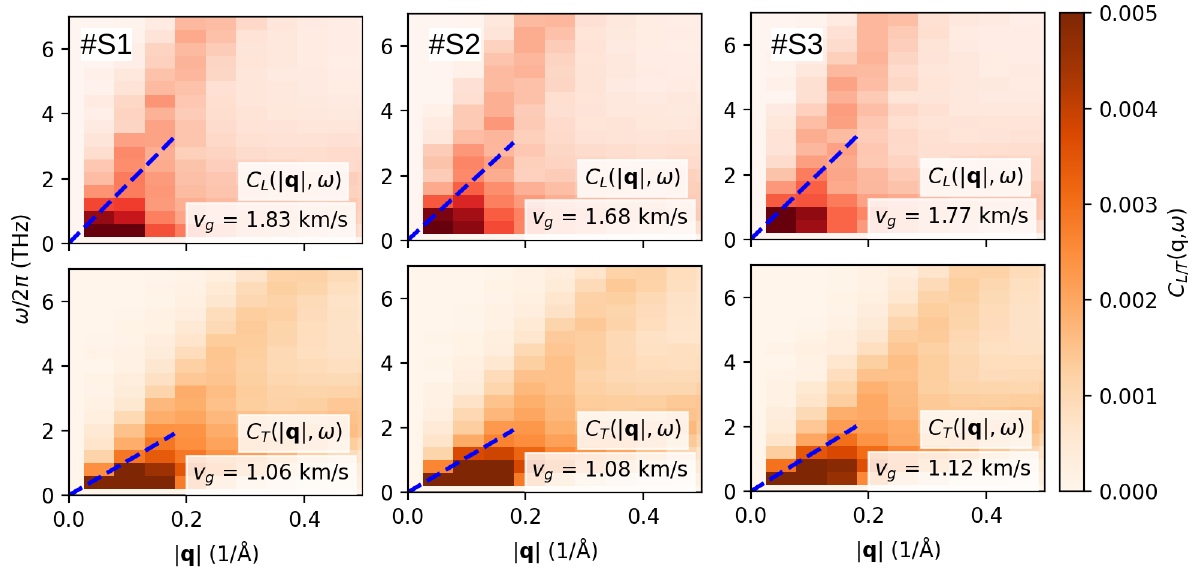}
\caption{
   Longitudinal (top $C_{\rm L})$) and transverse (bottom $C_{\rm T}$) current correlation functions as a function of wave vector ${q}$ and vibrational frequency $\omega/2\pi$ as calculated from three independent trajectories (\#S1-3) for 0.5 g cm$^{-3}$ NP-C. $10,000$ frames are extracted at equal intervals of $5$ fs from a 50-ps NVE-MD run for each trajectory. Group velocity ($v_{\rm g}$) at low frequencies is derived from the ratio of $\omega/2\pi$ to $q$ by a linear fitting in blue.}
\label{figs:NP-C-ccf-0.5}
\end{figure}

\begin{figure}
\addcontentsline{toc}{subsection}{\autoref{figs:NP-C-ccf-1}  Current correlation functions for $1$ g cm$^{-3}$ NP-C}
\centering
\includegraphics[width=1\columnwidth]{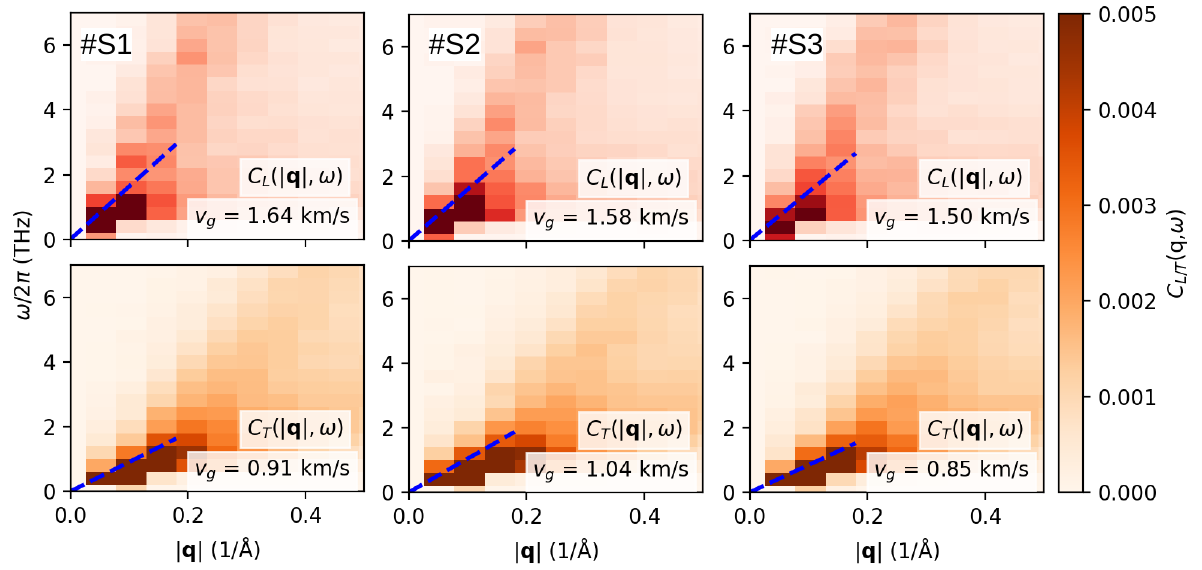}
\caption{Longitudinal (top $C_{\rm L})$) and transverse (bottom $C_{\rm T}$) current correlation functions as a function of wave vector ${q}$ and vibrational frequency $\omega/2\pi$ as calculated from three independent trajectories  (\#S1-3) for 1 g cm$^{-3}$ NP-C. $10,000$ frames are extracted at equal intervals of $5$ fs from a 50-ps NVE-MD run for each trajectory. Group velocity ($v_{\rm g}$) at low frequencies is derived from the ratio of $\omega/2\pi$ to $q$ by a linear fitting in blue.}
\label{figs:NP-C-ccf-1}
\end{figure}

\begin{figure}
\addcontentsline{toc}{subsection}{\autoref{figs:NP-C-ccf-1.5}  Current correlation functions for $1.5$ g cm$^{-3}$ NP-C}
\centering
\includegraphics[width=1\columnwidth]{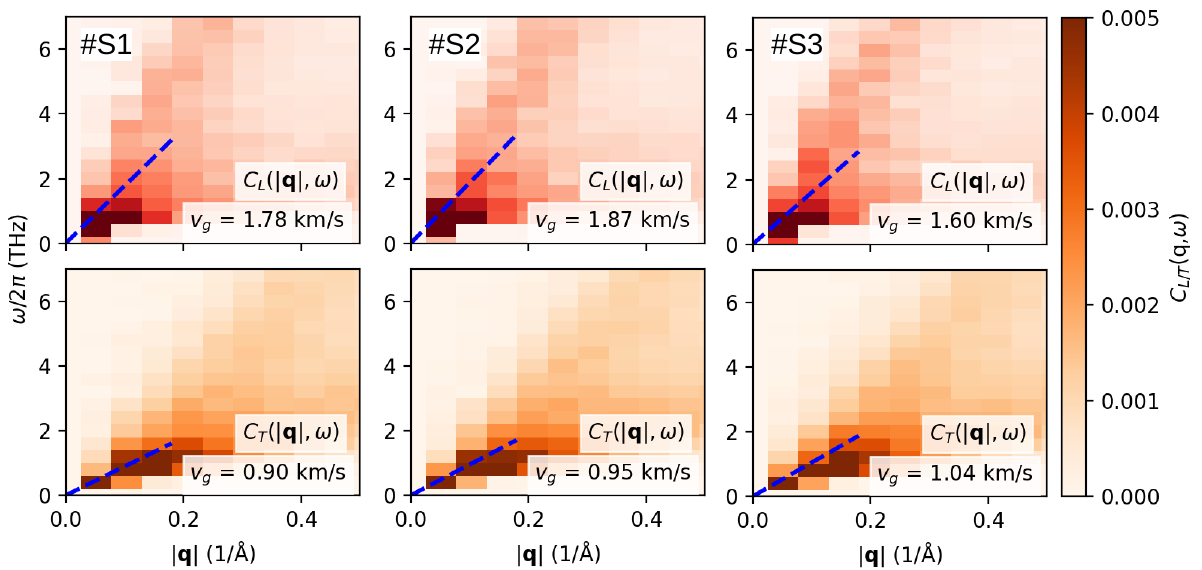}
\caption{Longitudinal (top $C_{\rm L})$) and transverse (bottom $C_{\rm T}$) current correlation functions as a function of wave vector ${q}$ and vibrational frequency $\omega/2\pi$ as calculated from three independent trajectories (\#S1-3) for 1.5 g cm$^{-3}$ NP-C. $10,000$ frames are extracted at equal intervals of $5$ fs from a 50-ps NVE-MD run for each trajectory. Group velocity ($v_{\rm g}$) at low frequencies is derived from the ratio of $\omega/2\pi$ to $q$ by a linear fitting in blue.}
\label{figs:NP-C-ccf-1.5}
\end{figure}

\begin{figure}
\addcontentsline{toc}{subsection}{\autoref{figs:aC-C-ccf-1.5} Current correlation functions for $1.5$ g cm$^{-3}$ a-C}
\centering
\includegraphics[width=1\columnwidth]{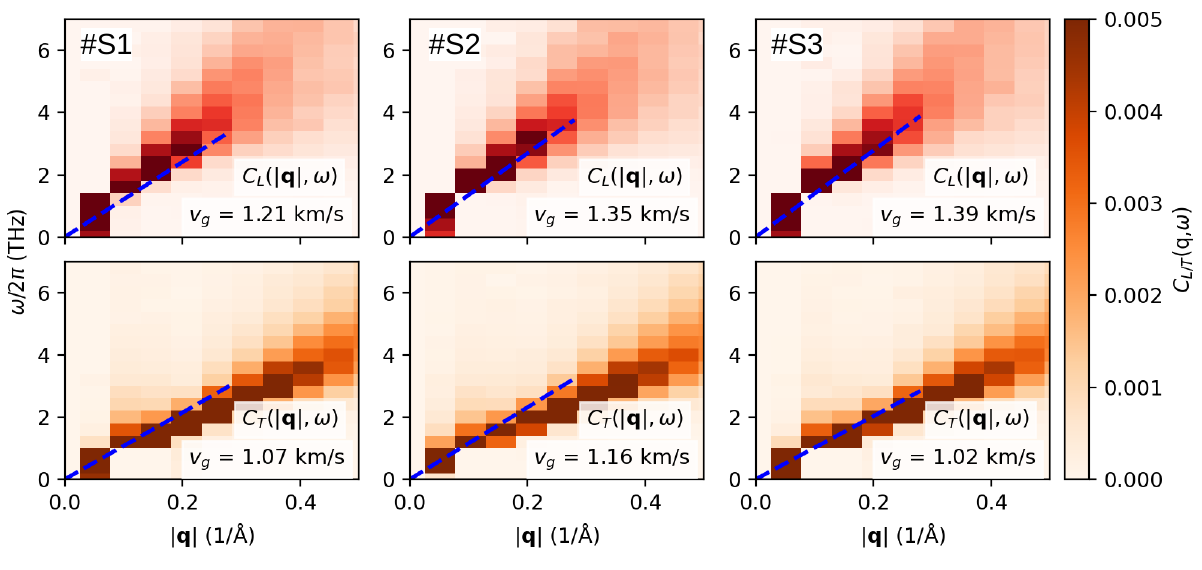}
\caption{
   Longitudinal (top $C_{\rm L})$) and transverse (bottom $C_{\rm T}$) current correlation functions as a function of wave vector ${q}$ and vibrational frequency $\omega/2\pi$ as calculated from three independent trajectories (\#S1-3) for 1.5 g cm$^{-3}$ a-C. $10,000$ frames are extracted at equal intervals of $5$ fs from a 50-ps NVE-MD run for each trajectory. Group velocity ($v_{\rm g}$) at low frequencies is derived from the ratio of $\omega/2\pi$ to $q$ by a linear fitting in blue.}
\label{figs:aC-C-ccf-1.5}
\end{figure}

\begin{figure}
\addcontentsline{toc}{subsection}{\autoref{figs:aC-C-ccf-2.5} Current correlation functions for $2.5$ g cm$^{-3}$ a-C}
\centering
\includegraphics[width=1\columnwidth]{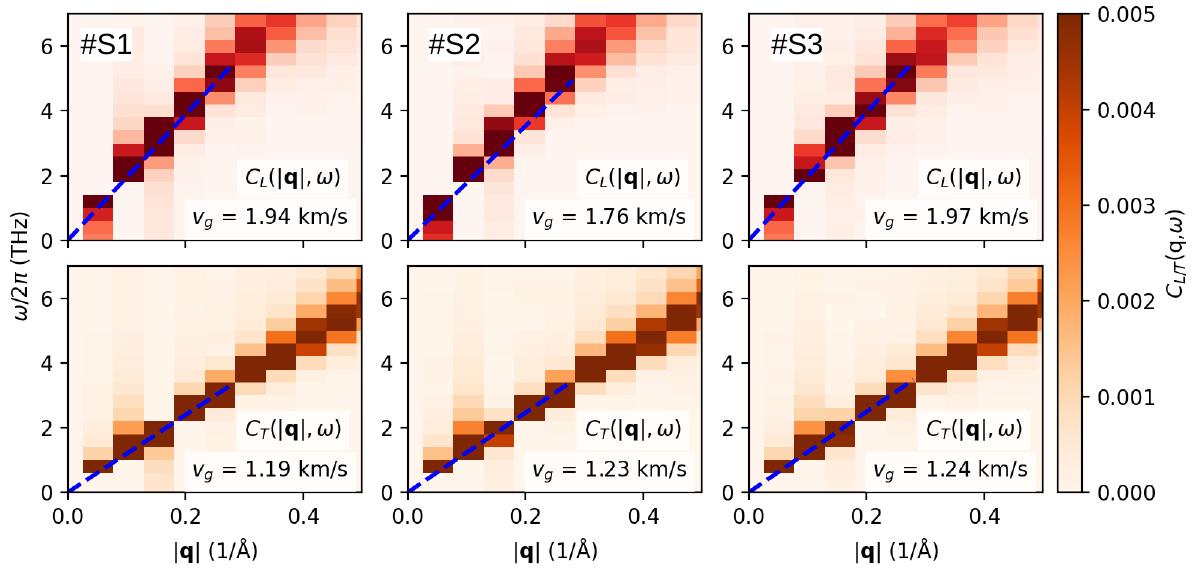}
\caption{ Longitudinal (top $C_{\rm L})$) and transverse (bottom $C_{\rm T}$) current correlation functions as a function of wave vector ${q}$ and vibrational frequency $\omega/2\pi$ as calculated from three independent trajectories (\#S1-3) for 2.5 g cm$^{-3}$ a-C. $10,000$ frames are extracted at equal intervals of $5$ fs from a 50-ps NVE-MD run for each trajectory. Group velocity ($v_{\rm g}$) at low frequencies is derived from the ratio of $\omega/2\pi$ to $q$ by a linear fitting in blue.}
\label{figs:aC-C-ccf-2.5}
\end{figure}

\begin{figure}
\addcontentsline{toc}{subsection}{\autoref{figs:aC-C-ccf-3.5} Current correlation functions for $3.5$ g cm$^{-3}$ a-C}
\centering
\includegraphics[width=1\columnwidth]{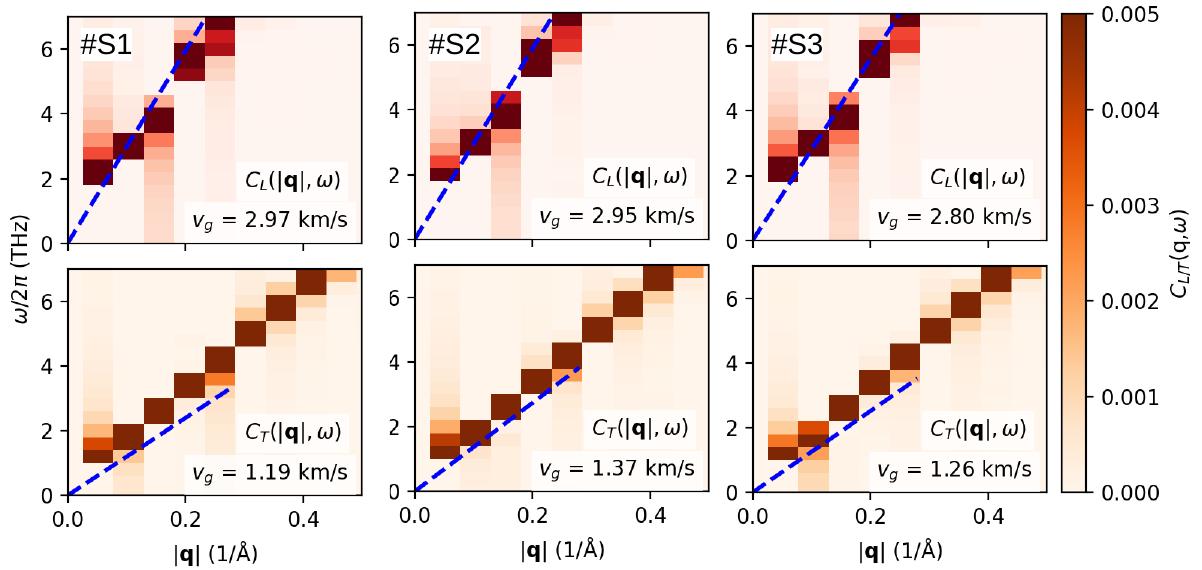}
\caption{Longitudinal (top $C_{\rm L})$) and transverse (bottom $C_{\rm T}$) current correlation functions as a function of wave vector ${q}$ and vibrational frequency $\omega/2\pi$ as calculated from three independent trajectories (\#S1-3) for 3.5 g cm$^{-3}$ a-C. $10,000$ frames are extracted at equal intervals of $5$ fs from a 50-ps NVE-MD run for each trajectory. Group velocity ($v_{\rm g}$) at low frequencies is derived from the ratio of $\omega/2\pi$ to $q$ by a linear fitting in blue.}
\label{figs:aC-C-ccf-3.5}
\end{figure}

\begin{figure}
\addcontentsline{toc}{subsection}{\autoref{fig:force_disorder} Force parities for disordered carbon}
    \centering
    \includegraphics[width=0.9\linewidth]{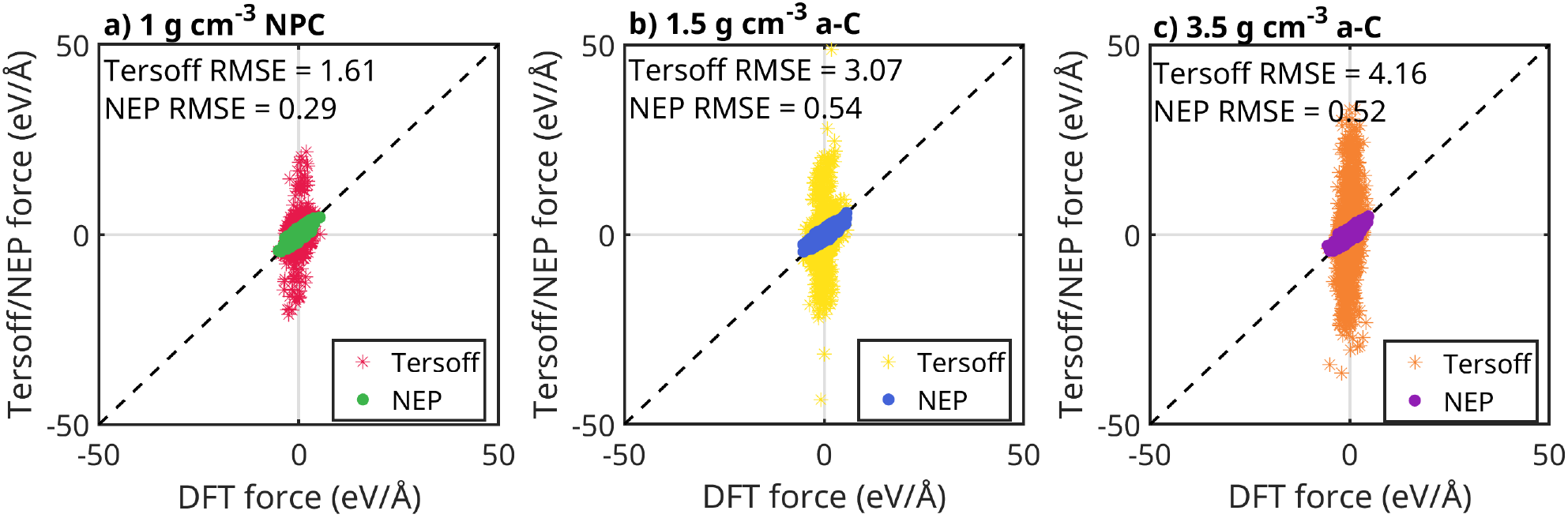}
    \caption{Force parities calculated using the NEP model and the Tersoff potential, compared to DFT-PBE reference forces for disordered carbon: (a) NP-C and (b) a-C. HNEMD runs were performed for $100$~ps on $1$~g~cm$^{-3}$ NP-C, and $1.5$~g~cm$^{-3}$ and $3.5$~g~cm$^{-3}$ a-C systems. From each run, $15$ equally spaced frames were sampled for atomic force calculations at each density. The corresponding RMSE values are shown in each panel.}
    \label{fig:force_disorder}
\end{figure}

\begin{figure}
\addcontentsline{toc}{subsection}{\autoref{fig:cumuKappa-crystal} Cumulative thermal conductivities as a function of production time for diamond}
    \centering
    \includegraphics[width=0.4\linewidth]{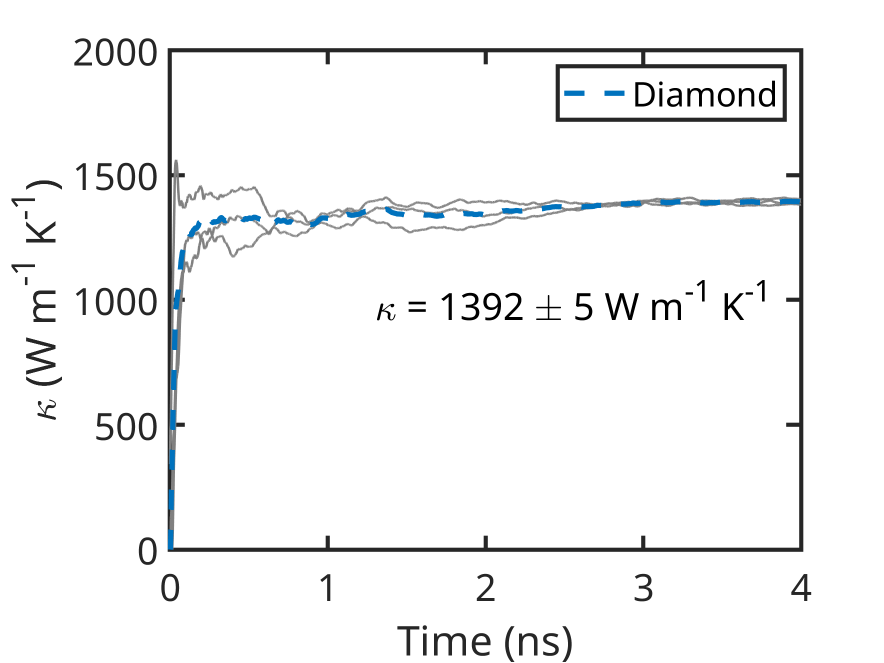}
    \caption{Cumulative thermal conductivities as a function of production time for diamond at $300$~K. Three independent HNEMD runs were performed, and the averaged thermal conductivity at the end of the production time is presented.}
    \label{fig:cumuKappa-crystal}
\end{figure}

The NEP model for carbon is trained using the hyperparameters listed in Note~\ref{ver:hyparameter}, and its trained accuracy is presented in Fig.~\ref{fig:figs_NEP-train}. Table~\ref{table:NPC-CN-density} summarizes the coordination fractions of sp, sp$^2$, and sp$^3$ bonds for NP-C samples with different densities. The size effect of thermal conductivity is illustrated in Fig.~\ref{figs:size-effect}. The cumulative $\kappa$ averaged over HNEMD production time is shown in Fig.~\ref{figs:kappa-time-NP-C} for NP-C samples and Fig.~\ref{figs:kappa-time-a-C} for a-C samples. Low-temperature NEMD spectral thermal conductance $G(\omega)$ is presented in Fig.~\ref{figs:Gw-vs-density_NP-C} and Fig.~\ref{figs:Gw-vs-density_a-C} for NP-C and a-C samples, respectively. Moreover, for 3 g cm$^{-3}$ a-C samples with different sp$^3$ contents, the cumulative HNEMD $\kappa(t)$ average is plotted in Fig.~\ref{figs:rho3-sp3-varing-kappa-vs-time}, $\kappa(\omega)$ is depicted in Fig.~\ref{figs:averaged_kw_sp3-varing}, and $G(\omega)$ is provided in Fig.~\ref{figs:Gw-vs-sp3}. Moreover, the current correlation functions for the low-frequency group velocities of NP-C and a-C samples are shown in Figs.~\ref{figs:NP-C-ccf-0.5}-\ref{figs:NP-C-ccf-1.5} and Figs.~\ref{figs:aC-C-ccf-1.5}-\ref{figs:aC-C-ccf-3.5}, respectively. Finally, the force parities for disordered carbon and cumulative thermal conductivities for diamond are presented in Fig.~\ref{fig:force_disorder} and Fig.~\ref{fig:cumuKappa-crystal}, respectively.

\tableofcontents{}

\clearpage
\newpage


\clearpage
\section{Supplemental Notes}

\notetitlelabel{snote:hyper}{Hyperparameters for NEP training}
The \verb"nep.in" input file for GPUMD reads:
\begin{verbatim} 
  version       3
  type          1 C
  cutoff        4.2 3.7
  n_max         8 6
  l_max         4 2
  basis_size    8 8
  neuron        100	
  lambda_1      0.05
  lambda_2      0.05
  lambda_e      1.0
  lambda_f      1.0
  lambda_v      0.1
  batch         6738
  population    50
  generation    250000
\end{verbatim}\label{ver:hyparameter}

\vfill
\section{\listtablename}

\begin{table}[h]
\addcontentsline{toc}{subsection}{\autoref{table:NPC-CN-density} Coordination fractions of NP-C }
    \centering
    \caption{Fractions of sp$^n$ ($1\leq n \leq 3$) motifs
     for NP-C with different densities (in units of g cm$^{-3}$).}
    \begin{tabular}{p{1.5cm} p{1cm} p{1cm} p{1cm} p{1cm} p{1cm} p{1cm}}
    \hline
    \hline
     & $0.3$ & $0.5$ & $0.75$ & $1$ & $1.25$ & $1.5$\\
    \hline
    sp (\%) & $1.38$  & $2.14$  & $2.04$ & $1.76$ & $1.59$ & $1.17$\\
    sp$^2$ (\%) & $98.56$  & $97.78$  & $97.82$ & $98.06$ & $98.15$ & $98.59$\\
    sp$^3$ (\%) & $0.03$  & $0.08$  & $0.13$ & $0.17$ & $0.26$ & $0.25$\\
     \hline
    \hline
    \end{tabular}
    \label{table:NPC-CN-density}
\end{table}

\clearpage
\section{\listfigurename}

\vfill

\clearpage

\vfill

\clearpage

\vfill

\clearpage

\vfill

\vfill

\clearpage

\vfill

\clearpage

\vfill

\clearpage

\vfill

\clearpage

\vfill

\phantomsection
\addcontentsline{toc}{section}{\listreferencename}

\bibliographystyle{myaps.bst}